\documentclass[11pt]{article}
\usepackage{natbib,epsf,rotating,latexsym} %,astron}
\bigskip

\newcommand{\be}{\begin{equation}}
\newcommand{\ee}{\end{equation}}
\newcommand{\bd}{\begin{displaymath}}
\newcommand{\ed}{\end{displaymath}}
\newcommand{\bea}{\begin{eqnarray}}
\newcommand{\eea}{\end{eqnarray}}

\newcommand{\eg}{{\it e.g.}}
\newcommand{\ie}{{\it i.e.}}

\newcommand{\m}{\,\hbox{m}}
\newcommand{\km}{\,\hbox{km}}
\newcommand{\second}{\,\hbox{s}}

\newcommand{\g}{\,\hbox{g}}

  \def\la{\mathrel{\mathchoice {\vcenter{\offinterlineskip\halign{\hfil
  $\displaystyle##$\hfil\cr<\cr\sim\cr}}}
  {\vcenter{\offinterlineskip\halign{\hfil$\textstyle##$\hfil\cr
  <\cr\sim\cr}}}
  {\vcenter{\offinterlineskip\halign{\hfil$\scriptstyle##$\hfil\cr
  <\cr\sim\cr}}}
  {\vcenter{\offinterlineskip\halign{\hfil$\scriptscriptstyle##$\hfil\cr
  <\cr\sim\cr}}}}}

\begin{document}

\sloppy
\parindent  0pt
\parskip    8pt

%-----------------------------------------------------------------------
% Title page
%-----------------------------------------------------------------------

\baselineskip 17pt
\vspace*{10mm}   

\begin{center}
\Huge
Impact-Generated Dust Clouds Surrounding the Galilean Moons
\end{center}

\vspace*{10mm}
\begin{center}
\Large
Harald~Kr\"uger$^{1}$,
Alexander V. Krivov$^{2,3}$,
Miodrag Srem\v{c}evi\'c$^{2}$, \\[4mm]
and Eberhard Gr\"un$^{1,4}$
\end{center}

\vspace*{7mm}
\begin{center}
\large
$^1$ 
Max-Planck-Institut f\"ur Kernphysik,\\
Postfach 103980, 69029 Heidelberg, Germany\\
E-Mail: Harald.Krueger@mpi-hd.mpg.de \\[6pt]
$^2$
Nonlinear Dynamics Group, Institute of Physics, University of Potsdam,\\
P.O. Box 601553, 14415 Potsdam, Germany\\
E-mail: krivov@agnld.uni-potsdam.de\\[6pt]
$^3$
On leave from: Astronomical Institute, St. Petersburg University,\\
Stary Peterhof, 198504 St. Petersburg, Russia\\
E-mail: krivov@astro.spbu.ru\\[6pt]
$^4$ Hawaii Institute of Geophysics and Planetology, University of Hawaii,\\
1680 East West Road, Honolulu, HI 96822, USA \\[6pt]
\end{center}

\vspace*{15mm}
\begin{tabular}{lr}
Manuscript pages: & 46\\
Figures:          & 16\\
Tables:           &  5\\
\end{tabular}

\vspace*{10mm}
\begin{center}
\Large
{\it Icarus}, in press

\thispagestyle{empty}
\vfill
\today\\
\end{center}

%-----------------------------------------------------------------------
% Second page
%-----------------------------------------------------------------------

\newpage
\vspace*{20mm}
\begin{center}
{\sl Proposed Running Head:}\\[1mm]
DUST CLOUDS OF THE GALILEAN MOONS
\end{center}

\vspace*{15mm}
\begin{center}
{\sl Corresponding author:}\\[1mm]
{\sf Harald Kr\"uger}\\[-1mm]
Max-Planck-Institut f\"ur Kernphysik\\[-1mm]
Postfach 103980\\[-1mm]
69029 Heidelberg, Germany\\
E-mail: Harald.Krueger@mpi-hd.mpg.de
\end{center}

\newpage

%-----------------------------------------------------------------------
% Abstract page
%-----------------------------------------------------------------------
\newpage
\vspace*{20mm}
\centerline{\large\bf Abstract}
\baselineskip 17pt

\medskip

%\begin{abstract}
Tenuous dust clouds of Jupiter's Galilean moons Io, Europa, Ganymede and Callisto 
have been detected with the in-situ dust detector on board the Galileo 
spacecraft. The majority of the dust particles have been sensed at 
altitudes below five radii of these lunar-sized satellites.
We identify 
the particles in the dust clouds surrounding the moons by their impact 
direction, impact velocity, and mass distribution.
Average particle sizes are between 0.5 and $\rm 1\,\mu m$, just above
%HK: changed:
the detector threshold, indicating a size distribution with decreasing numbers 
towards bigger particles.
Our results imply that the 
particles have been kicked up by hypervelocity impacts of micrometeoroids
onto the satellites' surfaces. The measured radial dust density profiles
are consistent with predictions by dynamical modeling
for satellite ejecta produced  by interplanetary impactors
(Krivov et al., {\em Planet. Sp. Sci.}, 2003, 51, 251--269), 
assuming yield, mass and velocity distributions of the ejecta from 
laboratory measurements. A comparison of 
all four Galilean moons (data for Ganymede published 
earlier; Kr\"uger et al., {\em Planet. Sp. Sci.}, 2000, 48, 1457--1471)
shows that the dust clouds of the three outer Galilean moons have very 
similar properties and are in good agreement with the model predictions
for solid ice-silicate surfaces. 
%HK: I kept this sentence contrary to Tadashi Mukai' suggestion because the
% title of the paper talks about all four galilean satellites. So we should
% mention Io in my opinion. 
%-AVK I agree. A compromise could probably be reached if you just extended the
%-AVK sentence to mention that the Io density was derived from 4
%-AVK individual impacts only ("a word of caution"). It's completely up to
%-AVK you though.
The dust density 
in the vicinity of Io, however, is more than an order of 
magnitude lower than expected from theory. This may be due
to a softer, fluffier surface of Io (volcanic
deposits) as compared to the other moons. 
%HK: changed: 
%-AVK Not the altitude! Distance from satelilite's center!
%-AVK The log-log slope of the dust number density vs. altitude 
%-AVK profiles of the clouds ranges
%-AVK between --1.3 and --2.8.
%HK-: I changed the Callisto slope (use only C3 and C10 now). 
The log-log slope of the dust number density in the clouds vs. distance from the
satellite center ranges
between --1.6 and --2.8.
Appreciable variations of number densities obtained from 
individual flybys with varying geometry, especially at Callisto, 
are found. These might be
indicative of leading-trailing asymmetries of the clouds due to the motion
of the moons with respect to the field of impactors.
%\end{abstract}

\medskip

\vspace*{2cm}
{\bf Keywords:}\\
\hspace*{3cm}dust\\
\hspace*{3cm}satellites of Jupiter\\
\hspace*{3cm}planetary rings

%=========================================================================

\newpage

\sloppy

\section{Introduction}

All celestial bodies without atmospheres are permanently exposed to
bombardment by hypervelocity micrometeoroids which knock-off
secondary ejecta dust particles from the surfaces of these bodies.
Impact ejection of dust particles has been suggested as 
the main process for maintaining dusty planetary rings like the 
Jovian rings
%%%AVK
%%%AVK Throughout the paper, I have replaced a \cite command
%%%AVK with correct commands of the natbib package:
%%%AVK     \citep which produces a citation in the form (Author, Year)
%%%AVK and \citet which produces Author (Year)
%%%AVK
\citep{morfill1980b,horanyi1996,ockert-bell1999,burns1999},
%%%AVK why don't us give more examples.... appropriate refs added
%%%AVK in paper.bbl (pls check)
%%%AVK Another point: colwell1993b did NOT consider E ring
%%%AVK and Saturn's E ring \citep{horanyi1992,colwell1993b,hamilton1994}.
Saturn's E ring
\citep{horanyi1992,hamilton1994}
as well as putative dust belts of Mars
\citep{soter1971,krivov1997,krivov1999} 
and Pluto \citep{thiessenhusen2002}.
With the in-situ dust detector on board the Galileo spacecraft 
\citep{gruen1992a}
a dust cloud formed by impact ejecta particles was for the first 
time detected surrounding Jupiter's moon Ganymede 
\citep{krueger1999d,krueger2000a}. 
The dust cloud was by far 
too tenuous to be detectable with remote sensing techniques.
Particles belonging to the Ganymede dust cloud were identified 
by their impact direction and impact speed and their 
sizes were mostly below $\rm 1\,\mu m$. 
The spatial distribution of the grains
as well as their size distribution were in agreement
with model predictions based on the impact ejection mechanism. 

The Galileo dust measurements can be treated as a natural 
impact experiment leading to the detection of the ejecta of 
hypervelocity impacts in space. They can give more insight
into the process of hypervelocity dust ejection, for
which the laboratory experiments on Earth still do not yield a
comprehensive picture. The measurements of the dust cloud at Ganymede 
stimulated the development of analytical models for  impact-generated
circumsatellite dust clouds not only for the Galilean moons but also 
for the Saturnian satellites \citep{krivov2003,sremcevic2003}. This is
especially important for the dust measurements to be collected 
at Saturn with the dust instrument \citep{srama2002} 
onboard the Cassini spacecraft beginning in 2004.

Since December 1995 Galileo has been on a bound orbit about Jupiter.    
The spacecraft had a
total of 32 targeted close flybys at all four Galilean moons: 7 encounters
with Io, 11 with Europa, 6 with Ganymede and 8 with Callisto. During many
of the encounters between 1995 and early 1999, the impact rate of 
dust grains showed a sharp peak within about half an hour centered at
closest approach to the moon 
\citep{gruen1996c,gruen1997b,gruen1998,krueger1998}.
These peaks indicated the existence of dust concentrations not only at
Ganymede but also in the close vicinities of Io, Europa and Callisto. During the
flybys at the Galilean moons after mid-1999, the spacecraft
orientation prevented the detection of dust particles close to 
the satellites. In November 2002, Galileo had the only
opportunity to in-situ measure dust in the close vicinity of
a fifth Jovian moon: Amalthea \citep{krueger2002}. 

In addition to dust clouds surrounding the Galilean moons, at least three
other populations of dust were detected by Galileo in the Jovian system
\citep{gruen1998}. Streams of 10-nanometer dust particles were detected 
throughout the Jovian magnetosphere and were recognizable even in 
interplanetary space out to 2~AU from Jupiter
\citep{gruen1993,gruen1994a}.
These dust grains originate from Io 
%HK: references added:
\citep{horanyi1993a,horanyi1993b,graps2000,krueger2003a}, their 
ultimate source probably being the most powerful of Io's volcanic plumes.
Bigger, micrometer-sized particles form a tenuous dust ring between the
Galilean moons and further away from Jupiter. Many of these particles
orbit Jupiter on prograde  orbits whereas a population on retrograde 
orbits exists as well 
\citep{colwell1998a,thiessenhusen2000,krivov2002a,krivov2002b}.
An overview of the Galileo dust measurements at Jupiter including the dust 
instrument itself can be found in \citet{krueger2003c}.

A detailed analysis of the dust grains detected at Galileo's four Ganymede
flybys in 1996 and 1997 has been published earlier
\citep{krueger1999d,krueger2000a}, showing that this Jovian moon is 
surrounded by a dust cloud formed by impact ejecta. Here, we 
analyze the dust impacts detected 
close to Io, Europa and Callisto, and compare our results with 
the measurements at Ganymede. Relevant physical properties of these 
moons are summarized in Table~\ref{phys_prop}.
In Section~\ref{sec_detection} we give a 
brief overview of the most important aspects of the Galileo dust instrument,
the Galileo spacecraft and the procedure  to identify
impacts of ejecta cloud particles in the Galileo dust data set. 
In Section~\ref{sec_data} we analyze
the dust detections at Io, Europa and Callisto.
In Section~\ref{sec_comp} we compare the properties of the dust
clouds of all four Galilean moons and check them against modeling.
Section~\ref{sec_conclusions} lists our conclusions.

\bigskip
\centerline{\fbox{\bf Insert Table \ref{phys_prop}}}
\medskip

%=========================================================================

\section{Dust impact detection}  \label{sec_detection}

\subsection{Galileo dust instrument} \label{sec_instrument}

The Galileo in-situ dust measurements at Ganymede provided the first
in-situ detection of an impact-generated dust cloud in space
\citep{krueger1999d}.
Processing of these measurements has been described in detail by 
\citet{krueger2000a}. Here we apply the same analysis
techniques to the dust measurements obtained in the close vicinities 
of Europa, Callisto and Io. Descriptions of
the dust instrument, Galileo  spacecraft, data transmission etc. have
been published in previous papers.
In what follows we recall only the most 
important aspects and give references to earlier publications where necessary. 

\begin{figure}
\epsfxsize=0.7\hsize
\epsfbox{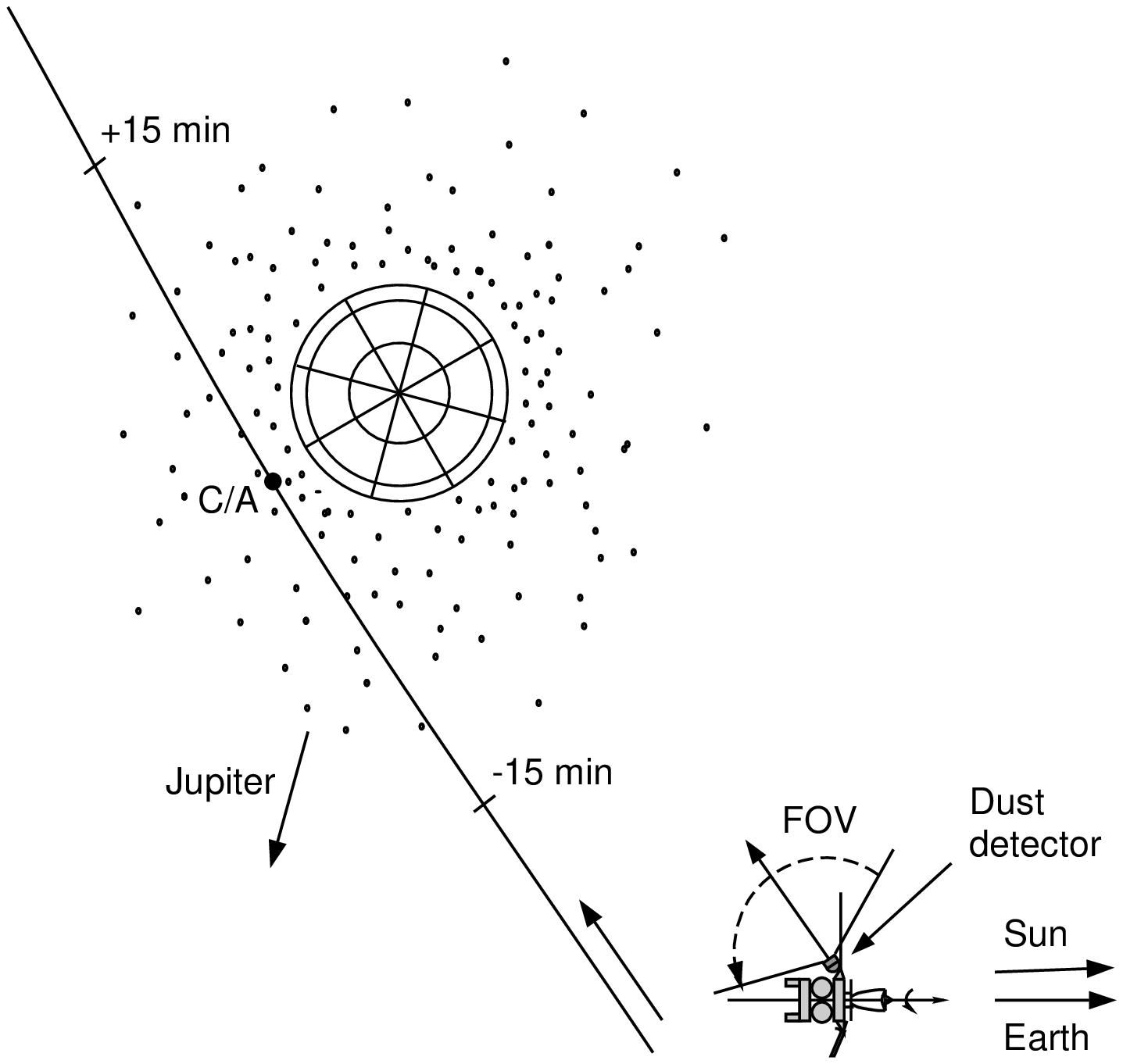}
        \caption{\label{geometry}
Galileo's trajectory and geometry of dust detection during the E4 
Europa flyby. The Galileo spacecraft is sketched in an orientation it
was in during the flyby (see text for details). The directions to
Jupiter, Earth and Sun are shown. C/A indicates 
closest approach to Europa, FOV the field of view of the
dust instrument. The orientation of the dust instrument shown corresponds to 
a rotation angle $\Theta = 270^{\circ}$.  At $90^{\circ}$ rotation angle 
it points in the opposite direction.
}
\end{figure}

Galileo is a dual-spinning spacecraft, with an antenna that points 
antiparallel to the positive spacecraft spin axis. During most of Galileo's
orbital mission about  Jupiter, the antenna pointed towards Earth. The
Galileo Dust Detector System (DDS), like its twin on-board Ulysses,
is a multi-coincidence impact ionization detector \citep{gruen1992a}
which measures submicrometer- and micrometer-sized dust impacts 
onto the detector target.
The dust instrument is mounted on the spinning section of Galileo 
and its sensor axis is offset by an angle of $60^{\circ}$ from 
the positive spin axis (Fig.~\ref{geometry}). 
Thus, during one spin revolution of the 
spacecraft, the detector scans the entire anti-Earth hemisphere, whereas
particles approaching from the Earth-ward direction remain undetectable.

For each dust grain hitting the detector target, three independent 
measurements of the impact-created plasma cloud are used to 
derive the impact speed $ v$ and  the mass $ m$ of 
the particle. The charge $ Q_{\rm I}$ released upon impact onto the target 
%HK: references added:
is described by the relation \citep{goeller1989,gruen1995a} 
\begin{equation} Q_{\rm I} \propto m \cdot v\,^{3.5} .
\label{equ1}
\end{equation} 
The calibrated speed range of the dust instrument is 
2 to $\rm 70\,km\second^{-1}$. The coincidence times of the three charge 
signals together with the charges themselves are used to classify each 
impact into one of four categories. Class~3 impacts have three 
charge signals, two are required for class~2 and class~1 events, 
and only one for class~0 \citep{baguhl1993b,gruen1995a,krueger1999a}.
Class 3 signals, our highest class, are real dust impacts and class 0 
events are noise. Class~1 and class~2 events are true 
dust impacts in interplanetary space  \citep{baguhl1993a,krueger1999a}.
However, in the Jovian system, 
within about 15$\rm R_J$ distance from Jupiter,
energetic particles from the Jovian plasma environment cause an 
enhanced noise rate in class~2 and the lower quality classes. 
By analysing the properties of the Io stream particles and 
comparing them with the noise events, the noise could be 
eliminated from the class~2 data \citep{krueger1999c}. All 
class~0 and class~1 events detected in the Jovian environment are 
usually classified as noise. 

Our noise identification scheme \citep{krueger1999c}, however, 
was derived for the Jovian dust 
%HK: changed:
stream particles and, hence, its applicability to other populations
of dust had to be verified.
Since Europa orbits Jupiter within the region where 
the high noise rates occurred, a slightly modified 
scheme has been developed for the ejecta grains detected in the close 
vicinity of Europa \citep[see ][their Table~4]{krueger2001a}.
It will be applied in this paper to remove noise events from the data sets 
obtained at the flybys at Europa and Io. For the Callisto data, no 
noise removal is necessary because Callisto orbits Jupiter outside the 
region where the high noise rates occurred. Noise removal was 
also not necessary in the earlier analysis of the Ganymede data 
\citep{krueger2000a}. It has to be noted that the noise removal technique 
uses statistical arguments and is applicable 
to large data sets only. Individual dust impacts may be erroneously 
classified as noise and vice versa. 

Galileo has a very low data transmission capability because its 
high-gain antenna did not open completely. For the dust measurements 
this means that the full set of parameters measured 
during a dust particle impact (spacecraft rotation angle, impact charges, 
charge rise times, etc.) could only be transmitted to Earth for a
limited number of impact events. During the close satellite flybys 
at the Galilean moons these limits were between one event per minute 
(record mode) and one event per 21 minutes (real time science mode) 
\citep{krueger2001a}. When event rates (\ie\ dust impacts plus noise) 
exceeded these numbers, the full set of parameters was 
transmitted to Earth for only a fraction of all detected events. All 
events, however, were always counted with one of 24 accumulators 
\citep{gruen1995a}. This way, the data can be corrected for
incomplete data transmission so that reliable impact rates can 
be determined for all satellite flybys \citep{krueger2000a}. 

Since its injection into an orbit about Jupiter, the Galileo spacecraft 
has been exposed to the harsh radiation environment of the Jovian 
magnetosphere with energetic particles of up to several MeV energies.
The Galileo Jupiter mission was extended three times so that 
the spacecraft was exposed to a total radiation dosage five times higher
than it was originally designed for. Especially high radiation dosages 
were acquired during orbit insertion in December 1995 and during several 
perijove passages after mid-1999 when Galileo's perijove 
distance from Jupiter was about $\rm 6\, R_J$.
It was anticipated that these high radiation levels would cause 
severe damages to the spacecraft electronics and the scientific instruments. 
Although degradation of the dust instrument was recognised in the dust 
data, no failure has occurred so far. The degradation effects
include -- amoung others -- a drop of the channeltron 
amplification, shifts of the measured instrument current, charge
rise times and amplitudes which reduced the sensitivity for dust 
impacts and noise events (Kr\"uger et al., in prep.). The
most important effect for our analysis here is shifts in the 
speed and mass calibration of the dust impacts.

\subsection{Impact direction}

The analysis of the dust measurements obtained at Ganymede showed that
the impact direction of the particles could be used as one important 
parameter to identify ejecta particles belonging to a 
dust cloud surrounding this moon \citep{krueger2000a}: 
in particular, the impact direction of the grains could be used to 
separate cloud particles from Jovian dust stream 
particles \citep{gruen1998}. As rotation angle, $\Theta$, we define the 
viewing direction of the dust sensor at the time of particle impact. 
During one spin revolution of the spacecraft, the rotation angle scans
through $360^{\circ}$. Rotation angles for the Galileo dust instrument,
however, are reported opposite to that of the actual spacecraft rotation
direction. This is done to easily compare Galileo results with the 
dust detector data taken on the Ulysses spacecraft,  which, unlike Galileo, 
has the opposite spin direction. Zero degrees of 
rotation angle is taken when the dust sensor points close to the 
ecliptic north direction. At rotation angles of $90^{\circ}$ and 
$270^{\circ}$
the sensor axis lies nearly in the ecliptic plane (which is close to 
Jupiter's equatorial plane). 

The dust instrument itself has a 
$140^{\circ}$ wide field of view (FOV). Dust particles which arrive within 
$10^{\circ}$ of the positive spin axis can be sensed at all rotation 
angles, while those that arrive at angles from $10^{\circ}$ to 
$130^{\circ}$ from the positive spin axis can only be sensed over a 
limited range of rotation angles. A sketch of the detection geometry
at close satellite flybys is shown in Fig.~\ref{geometry}.  

\subsection{Impact velocity}

Calibrated impact velocities are derived from the rise times of the impact
charge signals by an empirically derived algorithm \citep{gruen1995a}. 
The analysis of the dust impacts detected close to Ganymede showed that 
their average impact velocity onto the detector target was 
close to the encounter velocity of Galileo with this moon 
($\rm 8\,\km\second^{-1}$) \citep{krueger2000a}. It implied that the 
particles truly originated from Ganymede and that they belonged to a 
steady-state dust cloud surrounding this satellite. This good agreement 
of the measured mean impact velocity with the expected velocity also
showed that the calibration of the dust instrument is reliable in this
velocity range. 

Two statistical subsets of
particles could be separated: nearly all of the Ganymede particles
had calibrated velocities below $\rm 10\,\km\second^{-1}$, whereas
most of the stream particles had higher velocities. 
The calibrated impact velocity has been used as a 
parameter to separate both populations of dust at Galileo's 
G8 Ganymede flyby when the Jovian stream particles approached the 
dust detector from the direction towards Ganymede and particles 
belonging to Ganymede's steady-state dust cloud 
could not be identified by their impact direction alone.
The true impact velocities of stream
particles exceeded $\rm 200\,\km\second^{-1}$ \citep{zook1996} and were 
much faster than the velocity range of the
dust instrument calibrated in the laboratory ($\rm 70\,\km\second^{-1}$). 
Thus, the velocities for the stream particles derived from the instrument 
calibration significantly underestimate the true particle velocities.

\section{Data analysis}    \label{sec_data}

\subsection{Europa}

\subsubsection{Impact direction}  \label{sec_rot_eu}

The antenna of Galileo usually pointed towards Earth for data transmission.
This fixed the spin axis of the spacecraft so that the detector basically 
scanned the anti-Earth hemisphere.
In addition, due to the orbital motion of Jupiter about the Sun, the
geometry for dust detection with  the dust instrument gradually changed
with time, leading to the non-detectability of dust particles in the close
vicinity of the Galilean moons after mid-1999.
For this reason, ejecta particles were measurable during ten 
close Galileo flybys at Europa out of 11 flybys in total.
No data could be collected during two of these 
encounters due to spacecraft anomalies (safings) so that 
data sets from eight Europa flybys are available (E4, E6, E11, E12, E14, 
E15, E17, E19; see also Tab~\ref{tab_sat}). The labels of the encounters
are: the first letter of the satellite encountered by Galileo plus the 
number of Galileo's orbit about Jupiter.

\begin{figure}
\epsfxsize=0.8\hsize
\epsfbox{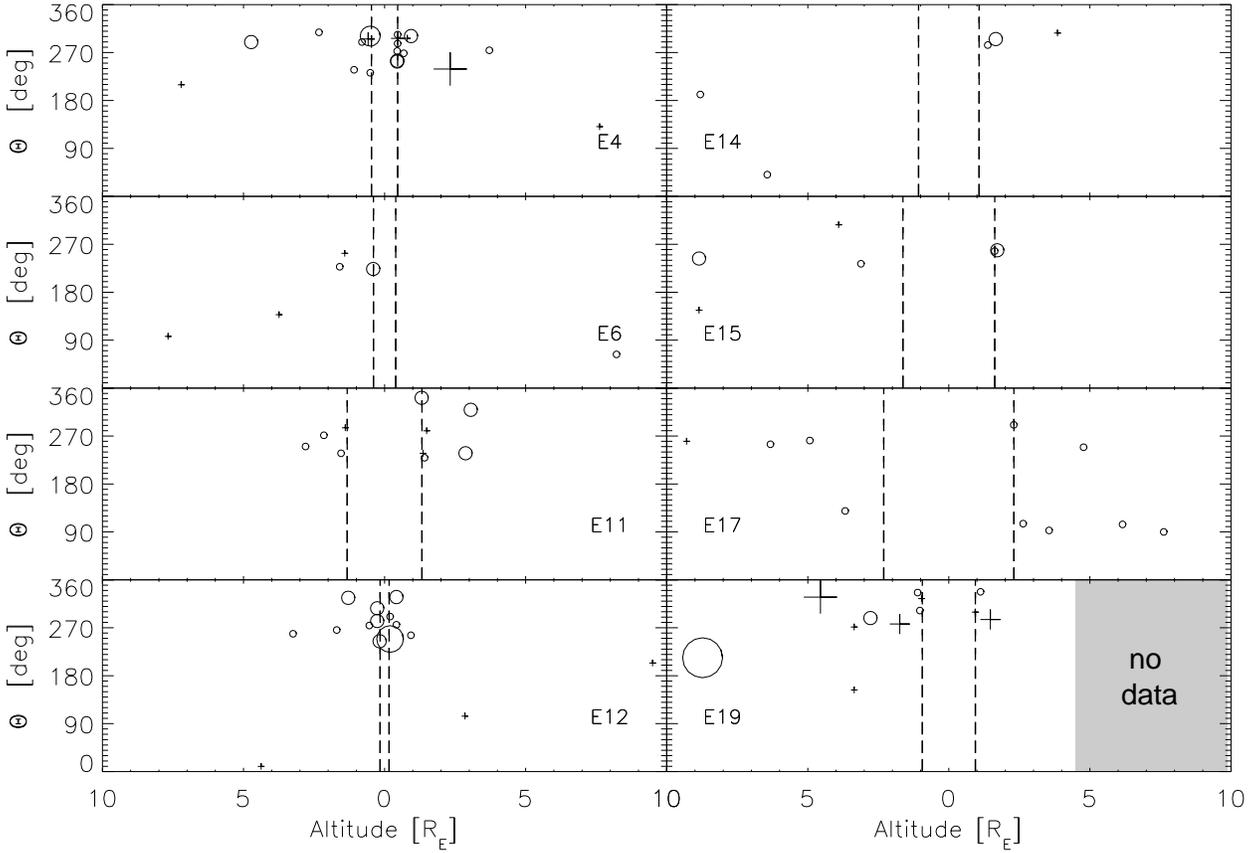}
        \caption{\label{rot_eu}
Sensor direction (rotation angle, $\Theta$) versus altitude of the Galileo 
spacecraft above the surface of Europa at the time of dust impact. 
Data are shown for all eight Europa encounters during which data were 
successfully collected (E4, E6, E11, E12, E14, E15, E17, E19). 
The radius of Europa, $\rm R_E$ is given in Table~\ref{phys_prop}.
The altitude range shown corresponds to a time interval of 2~h.
Each symbol indicates a dust particle impact, and 
the size of the symbol indicates the impact charge created by the 
particle ($  10^{-14}\,{\rm C} \leq Q_{\rm \,I} \leq 10^{-9}\,{\rm C}$).
Circles show particles with impact speeds below $\rm 10\,km\second^{-1}$ and
crosses show particles with higher speeds.
Galileo did not traverse the region between the vertical dashed lines.
In E19 no data were collected later than 19~min after closest approach 
(corresponding to 4.5 $\rm R_E$ altitude) because of a spacecraft anomaly.
Noise events have been removed \citep[][their Table~4]{krueger2001a}.
}
\end{figure}

In Fig.~\ref{rot_eu} we show the impact direction (rotation angle) of 
the dust particles detected within about 2~h around closest approach to 
Europa whose complete set of measured impact parameters has been transmitted 
to Earth. 
%Noise events have been removed by using the noise identification
%scheme of \citet{krueger2001a}. 
During most flybys,
particle impacts with $\rm 180^{\circ} < \Theta < 360^{\circ}$ 
were concentrated towards Europa. This is most obvious
during encounters E4, E11, E12 and E19. Most of these
impacts were detected at altitudes below $\rm 3\,R_E$
($\rm \,R_E$ is the Europa radius, see Table~\ref{phys_prop}).

To analyse the impact direction of the dust grains onto the detector
we assumed that the speed of dust relative to Europa in the vicinity of the moon
is low compared to Galileo's flyby speed.
Thus, the approach direction of the dust for an observer moving with 
the spacecraft is more or less parallel to the velocity vector of Europa
relative to the spacecraft (the so-called ram direction).
%HK-: changed
%Thus, an observer 'sitting' on the spacecraft 
%'sees' the particles approaching from a direction more or less parallel to the
%-AVK velocity vector of Europa (ram direction of the dust).
%velocity vector of the spacecraft (the so-called ram direction).
Since the orbital planes of Europa and Galileo about Jupiter
coincide to within a few degrees, such particles approached the detector
from a direction corresponding to $\rm \approx 270^{\circ}$ rotation angle 
during all eight encounters.
Rotation angles of about $\rm 90^{\circ}$ are opposite to the
direction towards Europa. With the sensor  field of view of 
$\rm 140^{\circ}$, particles detected with rotation angles 
$\rm 180^{\circ} < \Theta < 360^{\circ}$ are compatible with an origin 
from Europa itself. In the following we will call them Europa particles. 
This detection geometry is very similar to the one at the majority of the 
Ganymede flybys \citep{krueger2000a}.

\bigskip
\centerline{\fbox{\bf Insert Figure \ref{rot_eu}}}
\medskip

The direction from which the Jovian dust stream particles were observed varied 
during Galileo's path through the Jovian system: when Galileo approached the
inner Jovian system, rotation angles around $\rm 270^{\circ}$ were
observed. Between 1996 and early 1999 (the time span considered here) the rotation
angle shifted to $\rm 90^{\circ}$ shortly before Galileo's closest approach to 
Jupiter and the stream particles approached from this direction on the outbound 
portion of the spacecraft trajectory. 
Therefore, depending on when an individual satellite flyby occurred, stream particles 
approached the sensor from one or the other direction. In the cases of the
Europa flybys considered here, the stream particles approached from 
rotation angles $\rm 0^{\circ} < \Theta < 180^{\circ}$ (\ie\ opposite to
that of the Europa particles) or stream particle 
impacts had already ceased because of the unfavourable detection geometry.
It should be emphasized that during most of these eight Europa encounters 
more impacts were detected from the Europa direction than from the 
direction from which stream particles were to be expected.
Only between zero and two impacts from the direction of the stream particles
occurred during six encounters. Only encounters E6 and E17 showed the same number 
of impacts from $\rm 0^{\circ} < \Theta < 180^{\circ}$ (stream particles) 
as from the opposite direction (Europa particles). The statistics of 
particle detections is given in Table~\ref{tab_sat}.

\bigskip
\centerline{\fbox{\bf Insert Table~\ref{tab_sat}}}
\medskip

A total number of 64 Europa particles have been identified below 
$\rm 8\,R_E$ altitude during these eight encounters purely by their 
impact direction (Table~\ref{tab_sat}).
For our further analysis we use a cut-off altitude of 
$\rm 8\,R_E$ because this is close to the extension of Europa's Hill 
sphere (Table~\ref{phys_prop}). We can minimize
the potential contamination by particles belonging to other Jovian 
dust populations this way
(see Sect.~\ref{sec_rate_eu} for a discussion of their impact rates).
For instance, a cut-off altitude of $\rm 10\,R_E$ would increase the number 
of grains classified as Europa particles by only five. 

For some flybys, the numbers of 
identified Europa particles are lower limits to the true numbers of 
detected grains because the complete set of parameters measured upon impact 
could be transmitted to Earth for only a fraction of all impacts (column~9
of Table~\ref{tab_sat}). At E14, E15 and E19, however, the complete set
of parameters was transmitted for all impacts within 2~h around closest 
approach. 

In Table~\ref{tab_sat} (columns 7 and 8) we compare the number of Europa 
particles with the number of all events (dust plus noise) detected by the 
instrument below $\rm 8\,R_E$ and from a direction 
$\rm 180^{\circ} < \Theta < 360^{\circ}$. This shows that the noise
contribution to the total number of detected events in this altitude 
range is between 0 and 60\% with an average of 23\% (17 out of 81 events
are classified as noise). Although Europa orbits Jupiter in the region
where high noise rates 
occurred, the total number of noise events in the data set is relatively 
small. A plot similar to Fig.~\ref{rot_eu} but with all detected events 
(column~8 of Table~\ref{tab_sat}) also shows dust concentrations at the
Europa 
closest approaches so that the derived densities would also peak towards 
the satellite (Sect.~\ref{sec_rate_eu}). Hence, our conclusion that
the dust impacts are concentrated towards the surface of Europa  
does not depend upon the applied noise removal algorithm. We will come back to the noise 
problem in Sect.~\ref{sec_rate_eu} where we will determine the spatial distribution 
of dust surrounding Europa.
For our analysis of the Europa dust cloud we will 
use class~3 and noise-removed class~2 data. 
It should be noted that there is no physical difference between dust 
impacts categorised into class~2 and class~3.

\subsubsection{Impact velocity}   \label{sec_velocity_eu}

\begin{figure}
\epsfxsize=0.8\hsize
\epsfbox{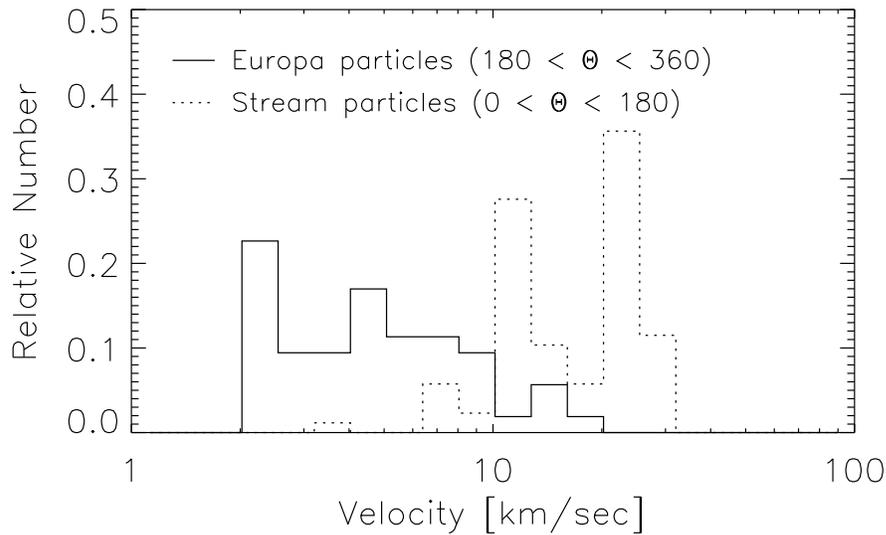}
        \caption{\label{velocities_eu}
Impact speeds derived from the instrument calibration for dust particles 
detected below $\rm 8\,R_E$ altitude at eight Europa encounters.
The solid line shows the 
distribution for Europa particles ($\rm 180^{\circ} \leq \Theta \leq 
360^{\circ}$) and the dotted line that for stream particles
($\rm 0^{\circ} \leq \Theta \leq 180^{\circ}$).
The mean impact speed is $\rm 5.5 \pm 3.5\,\km\second^{-1}$. Only 
particles with a velocity error factor $\rm VEF < 6$ \citep{gruen1995a} 
have been considered (53 Europa particles).
}
\end{figure}

In Fig.~\ref{velocities_eu} we show the velocity distribution of the 
Europa particles whose impact velocity has been determined with a
velocity error factor $\rm VEF < 6$ \citep{gruen1995a}. 53 particles
fulfill this criterion. 
During all eight encounters of Galileo with Europa, the flyby  
velocity was close to $\rm 6\,\km\second^{-1}$ (Table~\ref{tab_sat}),
well above the detection threshold of the dust instrument for micrometer-sized 
grains at $\rm 2\,km\second^{-1}$. We can separate two subsets of particles
from their velocity distribution, similar to the measurements at Ganymede: 
the Jovian stream particles with velocities typically above 
$\rm 10\,\km\second^{-1}$ and the slower Europa particles. 
The mean velocity of the 53 Europa particles in Fig.~\ref{velocities_eu} is 
$\rm 5.5 \pm 3.5 \,\km\second^{-1}$ ($\rm 1\, \sigma$). Given a typical 
uncertainty for an individual velocity measurement of a 
factor of two, this value is in good agreement with the 
velocity of Galileo relative to Europa.
%(about $\rm 6\,\km\second^{-1}$). 
%Spitze bei 2 km sec^{-1} diskutieren

\bigskip
\centerline{\fbox{\bf Insert Figure \ref{velocities_eu}}}
\medskip

The velocity measurements -- like the measurements at Ganymede -- are in 
agreement with dust particles belonging to a dust cloud of Europa.
They confirm  that the empirical velocity calibration of the dust instrument 
can be applied to the relatively slow ejecta particles, although the
calibration is wrong for the much smaller and faster Jovian dust stream 
particles. This result also confirms that the calibrated impact velocities 
can be used to identify particles belonging to a dust cloud when such grains 
cannot be identified by their impact direction. This will be applied to 
the data obtained in the vicinity of Callisto in 
Sect.~\ref{sec_ca_velocity}. It should be emphasized that
the two velocity distributions in Fig.~\ref{velocities_eu}
overlap, leading to some ambiguity in the identification of individual grains.
Hence, the velocity criterion can only be applied to a statistically large
data set.

In Fig.~\ref{rot_eu} we have marked particles according to their calibrated 
velocities: those with impact speeds below $\rm 10\,km\second^{-1}$ are shown as 
circles, faster grains as crosses. During seven Europa encounters of Galileo,
the majority of particles with $\rm 180^{\circ} \leq \Theta \leq 360^{\circ}$ had
impact speeds below $\rm 10\,\km\second^{-1}$, consistent with a particle
origin from Europa. Thus, for these seven flybys the identification of cloud 
particles with the velocity criterion is in agreement with the identificaton 
by the impact direction alone. Only the E19 encounter had 50\% of particles 
(5 out of 10) with higher calibrated impact speeds so that  
the majority of grains would be classified as stream particles and be rejected 
if the impact direction were not applicable as the main identification parameter.
Most of the particles detected at E19 with impact speeds above 
$\rm 10\,km\second^{-1}$ (4 out of 6 particles
with $\rm 0^{\circ} \leq \Theta \leq 360^{\circ}$), however, have a velocity 
error factor 
$\rm VEF > 6 $, which makes their speed calibration very uncertain, anyway. 
To summarize, the identification of Europa particles from their impact direction
and impact speed is quite reliable for all eight Europa encounters.
% Dies ist bei E19 nicht durch die Instrument-Alterung verursacht, da
% die Alterung einen Effekt in die andere Richtung erzeugen muesste.

\subsubsection{Impact rate and number density}    \label{sec_rate_eu}

\begin{figure}
\epsfxsize=0.6\hsize
\epsfbox{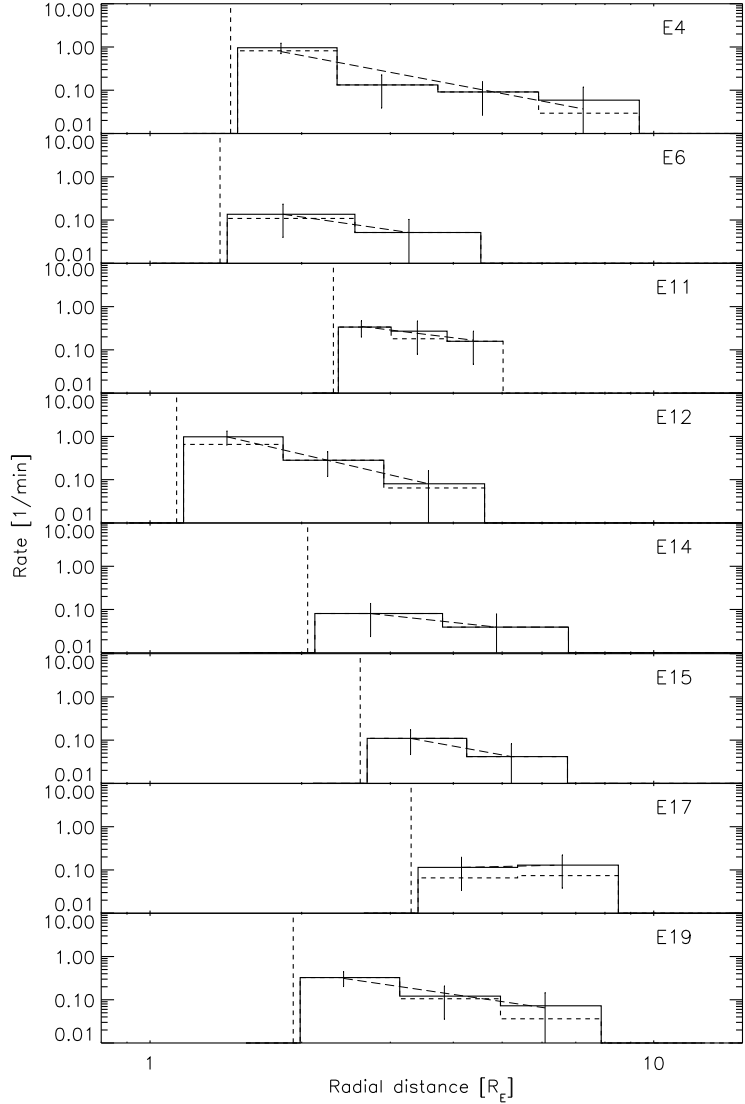}
        \caption{\label{rate_eu}
Impact rates of dust particles detected during eight Europa encounters.
The dotted histogram bins 
show the impact rates derived from the number of particles in each bin for which
their complete information has been transmitted to Earth. 
The solid histograms show the same rates, but corrected for incomplete data 
transmission. 
In cases where 
only a solid line is visible, no correction was necessary because
the complete information of all particle impacts is available in that bin.
The vertical dotted lines indicate the minimum altitude reached by Galileo
at closest approach.
Error bars denote the $\sqrt{N}$ statistical uncertainty, with $N$ being
the number of particles for which the complete information has been 
transmitted. Dashed lines are power law fits to the corrected impact
rate (Table~\ref{tab_sat}). In E17 data were collected with a low rate
of one instrument readout every 7~min which leads to a larger uncertainty
in the impact time. Therefore, the impact time of the
dust particles was shifted by --3.5~min to compensate for the reduced
time resolution. At E19, no data were collected beyond $\rm 4.5\,R_E$ after
Europa closest approach and the impact rate shown was multiplied by two 
outside this region to empirically correct for the missing data. 
}
\end{figure}

With 64 complete data sets of particles detected during eight Europa
encounters we can calculate the dust impact rate in 
the close vicinity of this moon (Fig.~\ref{rate_eu}). We have defined 
distance bins equally spaced on a logarithmic scale. Then we divided the 
number of particle impacts in each bin for which the complete set of 
measured impact parameters has been transmitted to Earth 
by the time Galileo has spent in that bin (dotted lines). 
Finally, to correct for incomplete data transmission 
(Sec.~\ref{sec_instrument}), we have multiplied the impact 
rate bin by bin with the ratio between the number 
of counted particles and the number of particles for which the complete 
data set has been transmitted. These corrected impact rates 
are shown as solid lines.

\bigskip
\centerline{\fbox{\bf Insert Figure \ref{rate_eu}}}
\medskip

Figure~\ref{rate_eu} shows that for those Europa encounters where
the number of detections is sufficiently large (at least ten particles;
E4, E11, E12 and E19) the impact rate clearly
increases towards Europa. This implies a concentration of dust 
particles at Europa. It was already obvious from Fig.~\ref{rot_eu} and
confirms the earlier results of a dust concentration at Ganymede 
\citep{krueger2000a}. On the other hand, at the 
remaining four encounters, the number of detections is so low that no 
statistically meaningful radial profile can be derived although the data 
are also compatible with particle concentrations at Europa. 
It should be noted that the 
correction for incomplete transmission is small in all bins and does 
not significantly affect the slopes of the power law fits. The slopes derived
with correction for incomplete transmission are in the range $-1.4$ to $-2.7$ 
(Table~\ref{tab_sat}). This is somewhat flatter than the most reliable
slopes obtained for Ganymede \citep{krueger2000a}. 

An important question arises:
is the slope of the dust distribution at Europa truly flatter than that
at Ganymede or is it an artefact caused by the measurement process? Two 
effects may cause a flattening of the slope:
1) a background of dust particles in jovicentric space;
and
2) incomplete removal of noise events by our noise removal technique.
Both would lead to an artificial flattening of the 
derived dust impact rate profile.

To analyze the first hypothesis  -- a background dust population -- we 
consider the complete data set of Galileo 
dust measurements in the Jovian system: the dust instrument has detected 
a number of micrometer-sized particles mostly in the region 
between the Galilean moons \citep{gruen1998}. At least two populations of 
grains can be distinguished: particles on bound prograde orbits about 
Jupiter and a population on retrograde orbits 
\citep{colwell1998a,colwell1998b,thiessenhusen2000,krivov2002a}. 
Depending on the detection geometry of the dust instrument during a 
specific orbit of Galileo about Jupiter, 
impact rates of particles from both populations taken together were  
up to six per day in the region of Europa
\citep[][their Fig.~10]{thiessenhusen2000}.
Considering that one Galileo passage through Europa's Hill sphere 
(Table~\ref{phys_prop}) lasted about 75~min, only one 
dust impact from these 
populations has to be expected every third Europa flyby. We therefore
conclude that a contamination by particles on jovicentric orbits 
is negligible for our analysis of the Europa dust cloud.
A potential contamination caused by 
such grains would be even smaller at Ganymede and Callisto because 
their number density decreases further away from Jupiter. 

In order to check the second possibility -- incomplete removal of 
noise events from the data set -- we use an alternative approach 
to calculate noise-free dust impact rates. The noise removal algorithm 
applied so far determines whether each individual class~2 event is 
most likely a true dust impact or a noise event. 
%HK: clarified, i hope:
Instead, we calculate the average noise rate 
measured with the dust instrument and subtract it from the total counted
rate to obtain the dust impact rate. We have first calculated the total event 
rate of dust impacts {\em plus noise} with the same technique as before, 
\ie\ from the complete data sets, without applying our noise-removal
scheme. This gives somewhat flatter power law 
slopes than those derived for the noise-removed data set 
(between $-1$ and $-2$).
We have then calculated the fraction of noise events in the class~2
accumulator data with our noise identification scheme during a one-day 
interval centered around each Europa flyby and 
calculated the rate of noise events in the counter data by multiplying 
the total counted rate with the fraction of noise events.
%HK changed text ends here.
Typical noise rates 
are between 0.04 and 0.1 per minute. We have then subtracted 
this noise rate from the total event rate obtained from the 
entire data sets. The resulting
radial density profiles have power law slopes between $-2$ and $-4$. 
%-AVK The theoretically expected value is $-2.5$ \citep{krivov2003}. 
%*MS* The theoretically expected value is about $-2.5$ \citep{krivov2003}. 
%*MS* 
%*MS* Colwell is right that the story with power-law slopes is not so simple, 
%*MS* and I propose the following slight expansion of the text.
The theoretically expected value is about $-2.5$, or more precisely 
it is steeper than $-2.5$ for $r\la 5\,R_{\rm sat}$ (bound grains dominate)
and it becomes flatter, between $-2.5$ and $-2$
(escaping grains dominate), farther out \citep{krivov2003}. 
%*MS* - end of changed text -
Thus, we conclude that the observed radial
density profiles which are somewhat flatter than the expected 
values may be due to incomplete noise removal.

%\bigskip
%\centerline{\fbox{\bf Insert Table~\ref{tab_noise_eu}}}
%\medskip

We do not investigate variations of the slopes between 
individual encounters because of the large statistical 
uncertainties and the potential unrecognised noise contamination 
of the data. Spatial variations with respect to the flyby 
position relative to the satellite will be addressed in a 
future investigation.

\begin{figure}
\epsfxsize=0.6\hsize
\epsfbox{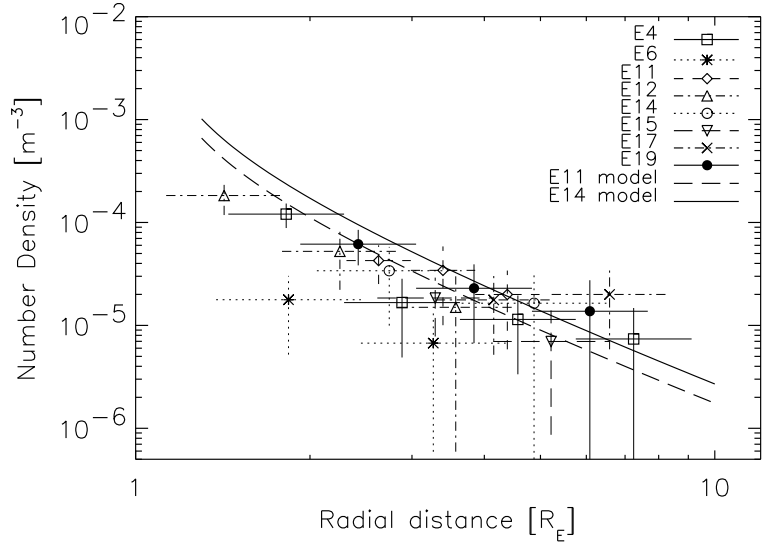}
        \caption{
\label{num_dens_eu}
The number density of dust as a function of distance from the center
of Europa, derived from the data (symbols with error bars) and
%-AVK predicted by the model \citep{krivov2003}.
predicted by the model \citep[][lines]{krivov2003}.
Horizontal bars for the data symbols indicate distance bins
which were used in processing the data (see text for details),
whereas vertical ones reflect $\sqrt{N}$ errors due to a limited
number of impacts.
%Two models are shown for different flyby speeds of Galileo.
%Since dust impacts are measured above a certain speed-dependent detection threshold
%of the instrument, different flyby speeds should give different dust number densities
%which has been taken into account in the model.
%HK: Josh Colwell did not understand the reason for the two model curves. 
%HK: Sasha, I tried a brief explanation, please check above. May be, this should
%HK: also be discussed in the text.
%-AVK Well explained! Maybe, a slightly extended version is even better:
Since dust impacts are measured above a certain speed-dependent detection threshold
of the instrument, different flyby speeds imply different minimum mass of the detected
grains and therefore should give different dust number densities.
This has been taken into account in the model.
Two model curves are shown: for the slowest (E11) and the fastest flyby (E14).
}
\end{figure}

\bigskip
\centerline{\fbox{\bf Insert Figure \ref{num_dens_eu}}}
\medskip

With the impact rate profiles derived for the individual
Europa encounters we can now calculate the spatial density
of dust in the environment of this satellite. We first 
divide bin by bin the impact rate by the spin-averaged
detector area to obtain fluxes ($\rm m^{-2}\,s^{-1}$). 
Then we divide these fluxes by the mean impact velocity
(spacecraft velocity relative to the moon) for a given
flyby. This results in mean number densities ($\rm m^{-3}$)
in the various distance bins.
Note that the slope of the number density is the same as that
of the impact rate, because both the spin-averaged detector area
and the mean impact velocity are assumed to be constant
(independent of distance) for any individual flyby.
The result is shown in 
Fig.~\ref{num_dens_eu}. The number densities show a 
clear increase towards Europa and the average slope is
$\rm -2.02\pm 0.63$. It
is remarkable that the variation in the derived number
densities from encounter to encounter is relatively
small. Since the closest approaches of Galileo at Europa
occurred at different longitudes and latitudes of Europa, 
it indicates that the dust distribution around this moon
does not show strong spatial or temporal variations.

%HK: I have added this paragraph in response to Tadashi's comments
%HK: no 5 and 6. May be, it should be more detailed and we should
%HK: give references?? Please check.
%-AVK Looks fine! Just one question (sorry for my ignorance):
%-AVK do you make a difference between "geysers" and "volcanoes"?
We have also checked alternatives for the origin of the
dust impacts detected in the close vicinity of the Galilean 
satellites other than the impact-ejection mechanism \citep{krueger2000a}.
Gravitational capture of the grains by the satellites can be 
dynamically ruled out. Electromagnetic interactions seem to be
too weak, in particular at Europa and Callisto which do not have 
their own magnetic fields. Although the geysers on Io are the most 
likely source for the Jovian dust stream particles, no geyser
activity has been observed on the other Galilean moons. 
The most plausible explanation for the origin of the dust grains was 
continuous bombardment of the satellites by interplanetary 
micrometeoroids.
%%%-AVK The rest of this paragraph, written obviously in
%%%-AVK response to Tadashi's comment (5), should probably be moved
%%%-AVK to the very beginning of this section 3.1.3.
%%%-AVK And indeed, already in the second sentence of this section
%%%-AVK you introduce "distance bins", which implicitly assumes
%%%-AVK distance to be the only spatial variable, which in turn
%%%-AVK assumes an approximate spherical symmetry! What do you think?
%*MS* 
%*MS* Maybe it could stay here as well, the number of impact events
%*MS* is low, so there is no way to introduce additional variables
%*MS* for binning. At the beginning it may be confusing for the reader.
%HK-: I prefer to keep it here.

In this work we assume spherical symmetry of the 
clouds which is supported by our measurements: passages of Galileo
at different latitudes and longitudes of the moons did not reveal
strong asymmetries in the dust distribution, except, possibly, 
at Callisto (see Sect.~\ref{sec_rate_ca}). This implies a 
spherical structure of the dust distribution surrounding the 
satellites. Of course, it does not rule out the existence of 
asymmetries in the dust density which will be investigated 
in the future \citep{sremcevic2003}.
%HK: end of new text here

We now look at the number density profile expected from theory.
\citet{krivov2003} developed a model of a spherically
symmetric, stationary dust cloud around a satellite, maintained by
impacts of interplanetary micrometeoroids. To the first approximation,
the number density of dust grains ejected into
ballistic orbits, which dominate the cloud at distances of several satellite
radii, is
\begin{equation}
  n_{\mathrm{bound}}(x) \propto  x^{-5/2},
  \label{bound_cal}
\end{equation}
where $x\equiv r/R_{\mathrm{sat}}$ is the distance measured in satellite radii.
The contribution of escaping grains into the cloud is somewhat shallower:
\begin{equation}
  n_{\mathrm{unbound}}(x) \propto  x^{-2}
  \label{unbound_cal}
\end{equation}
which slightly flattens the overall radial profile at larger
distances from the moon, closer to its Hill's sphere.

\citet{krivov2003} have also constructed an algorithm to
calculate the proportionality factors in
Eqs.~(\ref{bound_cal})--(\ref{unbound_cal}).
%essentially~--- the production rate of the grains (with masses exceeding the
%detector threshold) from the satellite surface.
The algorithm implies a chain of estimates: for the mass
flux and typical speed of projectiles, for gravitational focussing of impactors by
Jupiter, for the ejecta yield, ejecta mass and velocity distributions, etc.
The values of the model parameters (both assumed and derived) for Europa are given
in Table~\ref{tab_parms}.
Other parameters that have the same values for all Galilean satellites, are: 
slope of cumulative ejecta mass distribution $\alpha=0.83$, 
maximum mass of an ejected fragment $M_{\rm max}=10^{-5}\g$, 
opening angle of cone into which particles are ejected $\psi_0 = 90^\circ$.
For a detailed description of the parameters, the reader is referred to the original 
paper.
%-AVK I added the following stuff (Josh's comment 21)
We note that the model calculates number densities of particles with masses
above the detection threshold of the Galileo dust detector.
As the threshold is speed-dependent, the number densities are computed separately
for each Galileo flyby, and the results for specific flybys generally
differ even for the same moon and the same distance.
A strong advantage of this approach is that it enables direct
comparison of the number densities predicted by the model with those derived
from the measurements.
%-AVK New text ends here

Using this model for Europa, we obtained theoretical curves superimposed on the
data points in Fig.~\ref{num_dens_eu}.
A comparison between the number densities derived from the Galileo measurements
and those computed with the model will be given in Sect.~\ref{sec_comp}.

\bigskip
\centerline{\fbox{\bf Insert Table \ref{tab_parms}}}
\medskip

\subsubsection{Mass distribution}

The charge released by an impact of a dust particle onto
the detector target depends on the mass and the velocity of the 
grain (Equ.~\ref{equ1}).
In particular, to calculate the particle mass one has to know its 
impact velocity. The calibration of velocity and mass from the
measured charge rise times and charge amplitudes is usually
performed based on laboratory measurements obtained at a
dust accelerator.

\begin{figure}
\epsfxsize=0.6\hsize
\epsfbox{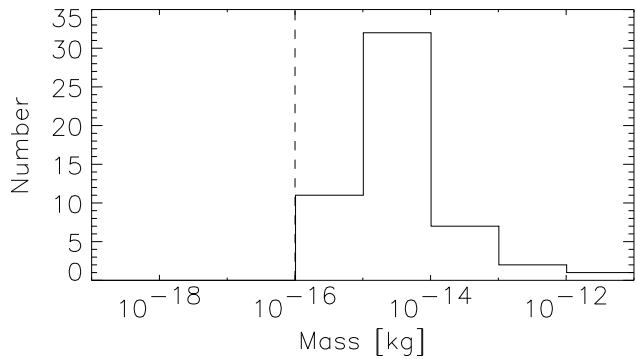}
\vspace{-8cm}

\epsfxsize=0.6\hsize
\epsfbox{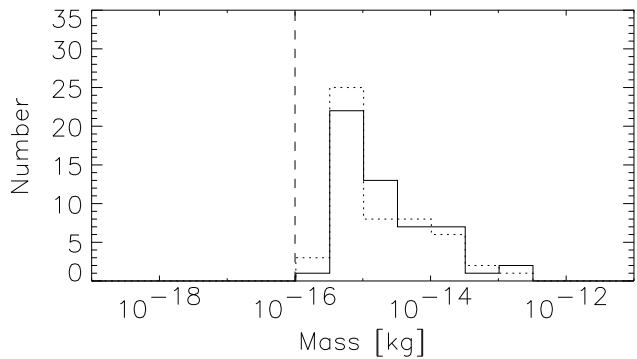}
        \caption{
\label{mass_hist_eu}
Mass distribution of the particles detected during eight Europa flybys.
The upper panel shows the distribution obtained by using the 
measured impact speeds derived from the instrument calibration. 
In the lower panel the speed of Galileo relative to Europa has been 
assumed as the impact velocity in order to calculate the particle mass.
The dotted histogram is without aging correction, the data for 
the solid histogram
has been corrected for aging of the dust detector electronics.
The vertical dashed lines indicate the detection threshold for 
particles which approach the detector with the velocity of
Galileo relative to Europa  (about $\rm 6 \km\second^{-1}$). 
Only the 53 particles with a velocity error factor 
$\rm VEF < 6$ \citep{gruen1995a} have been considered.
}
\end{figure}

In Fig.~\ref{mass_hist_eu} we show the mass distribution of the
particles from all eight Europa flybys for which the velocity
could be reliably determined ($\rm VEF < 6$; 53 particles in total). 
In the upper panel the complete instrument calibration has been
used to obtain particle speed and mass. With this method the 
uncertainty of the impact velocity is typically a factor of 2 and that 
of the mass is a factor of 10.

\bigskip
\centerline{\fbox{\bf Insert Figure \ref{mass_hist_eu}}}
\medskip

The dust detector has a velocity-dependent
detection threshold \citep{gruen1995a}. The threshold
for particles approaching with $\rm 6 \km\second^{-1}$
is shown as a dashed line. The mass distribution is incomplete
around this value.

The mass distribution is also affected by the low data transmission 
capability of Galileo and the data storage scheme in the instrument 
memory. As a result, nearly all data sets lost are in the lowest 
amplitude range AR1 which --- for particle velocities of 
about $\rm 6\, km\,s^{-1}$ --- corresponds to the mass
range below  $\rm \sim 3 \times 10^{-15}\,kg$. If we assume 
that the lost particles are equally distributed over the mass bins
below this value, the maximum of the mass distribution is 
artificially too low by less than a factor of 1.2. Thus, incomplete
data transmission does not significantly affect the mass distribution for 
Europa particles.

If the individual impact velocities of dust particles were known 
with a higher accuracy than the typical factor of 2 uncertainty from 
the instrument calibration, the uncertainty in the mass determination 
could be improved. The measured mean impact velocities of Europa particles 
are close to the velocity of Galileo relative to Europa during the 
individual encounters (Table~\ref{tab_sat}). We therefore 
assume the latter ones as the particles' impact velocities and 
show the recalculated particle mass in the lower panel of 
Fig.~\ref{mass_hist_eu}. The width of the mass distribution is 
significantly smaller than that derived from the calibrated impact 
velocities. 
This method has also been successfully applied to 
calculate the size distribution of Ganymede ejecta particles 
\citep{krueger2000a} and interstellar dust particles measured with 
Galileo and Ulysses \citep{landgraf2000a}. 

The mean mass of the Europa particles is 
$\rm 9.0 \times 10^{-15}\,kg$. Assuming spherical particles with a 
density of $\rm 1\,g\,cm^{-3}$ -- the density of water ice -- this 
corresponds to a particle radius of $ \rm \approx 1 \,\mu m$. 

Degradation of the dust instrument caused by the high radiation
dosages in the Jovian magnetosphere lead to shifts of the calibrated
masses and impact speeds of the dust particles. All data
collected after mid-1997 are affected by this shift, the later in 
the mission the data were collected, the stronger the shift. For our 
Europa measurements this means that masses are too low by a factor 
of about 1.5 beginning with the E11 encounter.  We have corrected 
the calibrated masses for these data sets and constructed a 
corrected mass distribution (solid histogram in the bottom panel of 
Fig.~\ref{mass_hist_eu}). Since we have taken the speed of 
Galileo relative to Europa as the impact speed of the particles, we 
need to correct the masses only. The shift in the velocity calibration 
caused by the instrument degradation does not affect this mass
distribution. The resulting mean mass of the Europa particles is
$\rm 1.3 \times 10^{-14}\,kg$. It should be emphasized that the 
shift in the velocity calibration does not affect the identification
of the Europa particles because the particles at the Europa
encounters have been identified by their impact direction alone, 
without using the impact speed as an additional criterion.

\subsection{Callisto}

\subsubsection{Impact direction}

\begin{figure}
\epsfxsize=0.8\hsize
\epsfbox{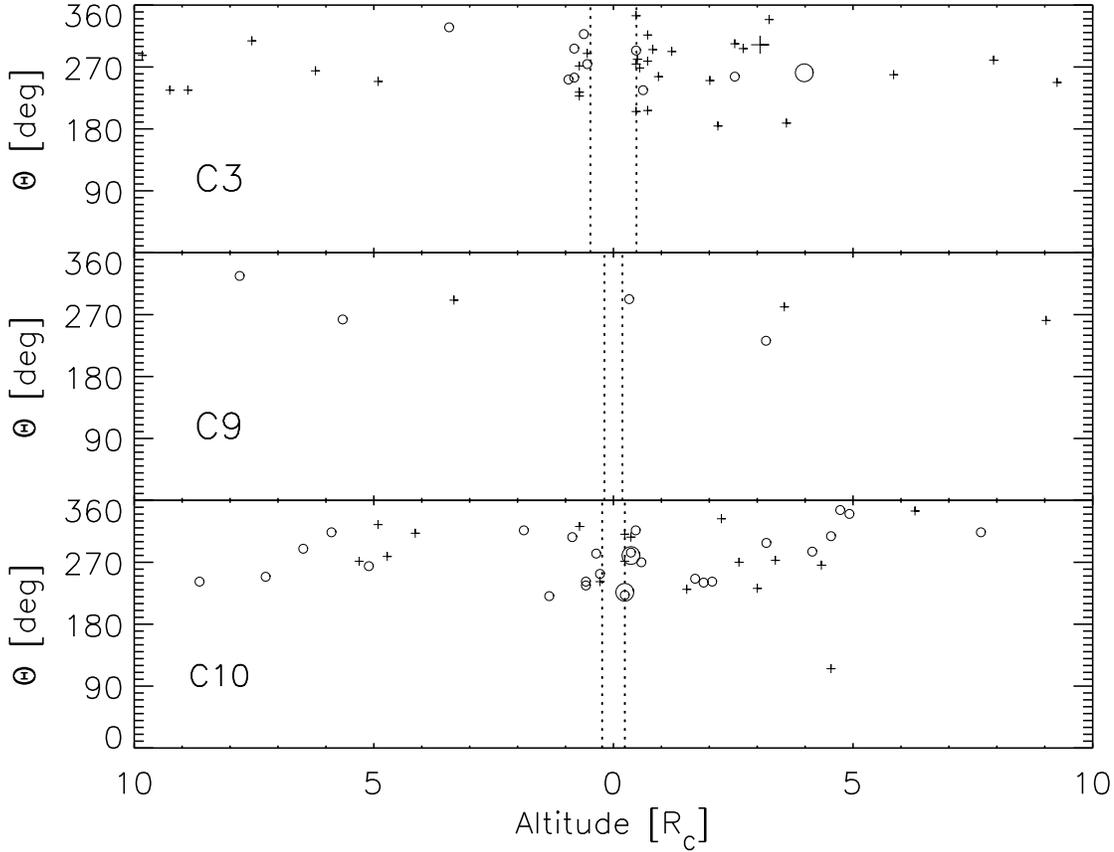}
        \caption{\label{rot_ca}
Sensor direction (rotation angle, $\Theta$) versus altitude of the Galileo
spacecraft above the surface of Callisto at the time of dust impact.
Data are shown for all three Callisto encounters at which the spacecraft 
orientation allowed the detection of cloud particles
(C3, C9, C10). The radius of Callisto, $\rm R_C$, is given in
Table~\ref{phys_prop}.
The altitude range shown corresponds to a time interval of 2 h.
Each symbol indicates a dust particle impact, and
the size of the symbol indicates the impact charge created by the
particle ($ \rm 10^{-14}\,C \leq Q_{\,I} \leq 10^{-13}\,C$).
Circles show particles with impact speeds below $\rm 10\,km\second^{-1}$ and
crosses show particles with higher speeds.
Galileo did not traverse the region between the vertical dashed lines.
}
\end{figure}

During Galileo's prime mission about Jupiter in 1996 and 1997 the
spacecraft had three close flybys at Callisto (C3, C9, C10). 
%%%AVK My dictionaries say, "allow for"~"take into account"
%%%HK: you are right.
%%%AVK The spacecraft orientation during these encounters allowed for the 
The spacecraft orientation during these encounters allowed the 
detection of ejecta cloud particles close to Callisto.
During all Callisto flybys after mid-1999 the spacecraft orientation
prevented the detection of ejecta cloud particles so that the
measurements at Callisto are restricted to these three encounters.
Unfortunately, all three of them occurred on the portion of the 
Galileo trajectory inbound to Jupiter where Jupiter stream 
particles and potential ejecta particles from Callisto approached 
the dust sensor from the same direction ($\rm 180^{\circ} \leq 
\Theta\leq 360^{\circ}$; Sect.~\ref{sec_rot_eu}). 
This is shown in Fig.~\ref{rot_ca}:
almost the entire number of dust impacts measured close to 
Callisto occurred from this direction. Thus, Callisto particles 
could not be uniquely identified by their impact direction alone, 
and we had to use the impact velocity as an additional criterion 
to identify them.

\bigskip
\centerline{\fbox{\bf Insert Figure \ref{rot_ca}}}
\medskip

The analysis of the ejecta particles detected at Europa 
(Sect.~\ref{sec_velocity_eu}) and Ganymede \citep{krueger2000a} showed
that -- on average -- stream particles and ejecta cloud particles
occupy different regimes in calibrated impact speed: cloud particles
have typical speeds below $\rm 10\,km\second^{-1}$ which are on average
very close to the encounter velocity of Galileo with the satellite, whereas 
stream particles have significantly higher calibrated speeds. This has been
successfully applied to identify Ganymede cloud particles from Galileo's 
G8 encounter at Ganymede. We apply the same velocity
criterion here to separate Callisto particles from the dust streams. 
The numbers of Callisto particles identified this way 
are listed in Table~\ref{tab_sat} for each orbit. The total number of 
Callisto particles from all three encounters is 35.

Only particles detected at altitude below $\rm 6\,R_C$ are 
considered for further analysis in order to minimize the contamination by 
stream particles. The analysis of the Europa and Ganymede data showed
that a few stream particles have calibrated 
velocities below $\rm 10\,\km\second^{-1}$ and would erroneously be 
classified as Callisto particles (Fig.~\ref{rot_ca}). Most cloud
particle impacts at Europa and Ganymede occurred below an altitude of 
about $\rm 6\,R_{\rm sat}$ (Fig.~\ref{rot_eu} and \citep{krueger2000a})
so that the inclusion of particles detected further 
away would increase the probability that the particles are actually
stream particles rather than dust cloud particles.
An apparent concentration of stream particle impacts within 
$\rm 3\,R_C$ altitude at C3 and within $\rm 5\,R_C$ at C10 
(Fig.~\ref{rot_ca}) is due to a higher data transmission rate of 
Galileo in these periods \citep{krueger2001a}.

Callisto orbits Jupiter outside the region within $\rm 15\,R_J$ where 
the high noise rates occurred. Thus, a potential noise contamination 
of the Callisto data is expected to be very low.
For our analysis of the Callisto dust cloud we will use class~3 and
class~2 data without noise removal. 

\subsubsection{Impact velocity}   \label{sec_ca_velocity}

\begin{figure}
\epsfxsize=0.7\hsize
\epsfbox{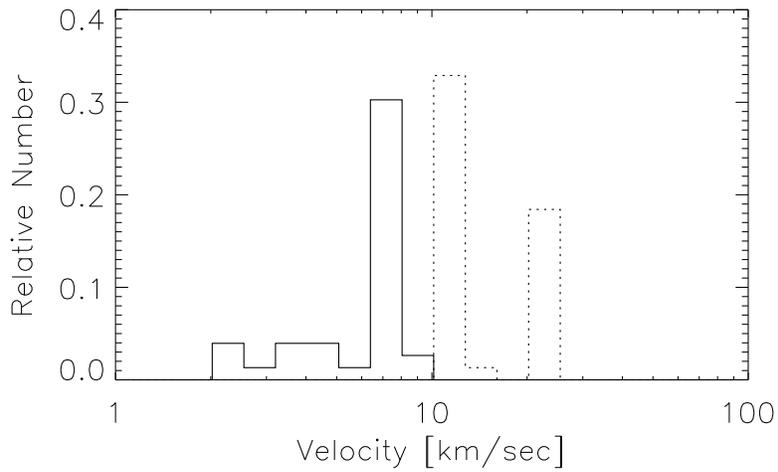}
        \caption{\label{velocities_ca}
Impact speeds derived from the instrument calibration for dust particles
detected at all three Callisto encounters
below $\rm 6\,R_C$ altitude with $\rm 180^{\circ} \leq \Theta \leq
360^{\circ}$ and 
calibrated impact velocities below $\rm 10\,km\second^{-1}$ (solid 
line) and above $\rm 10\,km\second^{-1}$ (dotted line).
%The average impact velocity is $\rm 6.4 \pm 2.1\,\km\second^{-1}$. 
Only
particles with a velocity error factor $\rm VEF < 6$ \citep{gruen1995a}
have been considered (35 particles with $\rm 10\,km\second^{-1}$).
}
\end{figure}

A total number of 35 Callisto particles have been identified by their
calibrated impact velocity below $\rm 10\,\km\second^{-1}$ and below 
an altitude of $\rm 6\,R_C$.
%For C3 (Fig.~\ref{rot_ca}) only a small fraction of particles with 
%$\rm 180^{\circ} \leq \Theta \leq 360^{\circ}$ have 
%calibrated impact velocities below $\rm 10\,\km\second^{-1}$ (10 
%out of 35, see  Table~\ref{tab_sat}). For C9 and C10 the fraction of slow
%particles increases (4 out of 7 for C9 and 27 out of 44 for C10,
%respectively). 
Their mean impact velocity is $\rm 6.4 \pm 2.1\,\km\second^{-1}$
($1\,\sigma$). 
This value is artificially too low because the velocity distribution
is cut off at $\rm 10\,\km\second^{-1}$. 
The average flyby speed of Galileo at Callisto was
$\rm 8.1\,\km\second^{-1}$ and both speeds agree within $1\,\sigma$. 
%The speed distributions
%of Europa and Ganymede particles also showed higher calibrated impact 
%speeds.
The velocity distribution of the Callisto particles is shown in 
Fig.~\ref{velocities_ca}.

%*MS* this reference is obsolete now
%*MS* \bigskip
%*MS* \centerline{\fbox{\bf Insert Table \ref{tab_sat}}}
%*MS* \medskip

\bigskip
\centerline{\fbox{\bf Insert Figure \ref{velocities_ca}}}
\medskip

Degradation of the dust instrument electronics does not significantly 
affect the impact velocities of the particles because most the 
Callisto encounters occurred relatively early during the Galileo mission.
In particular, the identification of Callisto particles via their impact
speed is not affected.

\subsubsection{Impact rate and number density}
   \label{sec_rate_ca}

\begin{figure}
\epsfxsize=0.9\hsize
\epsfbox{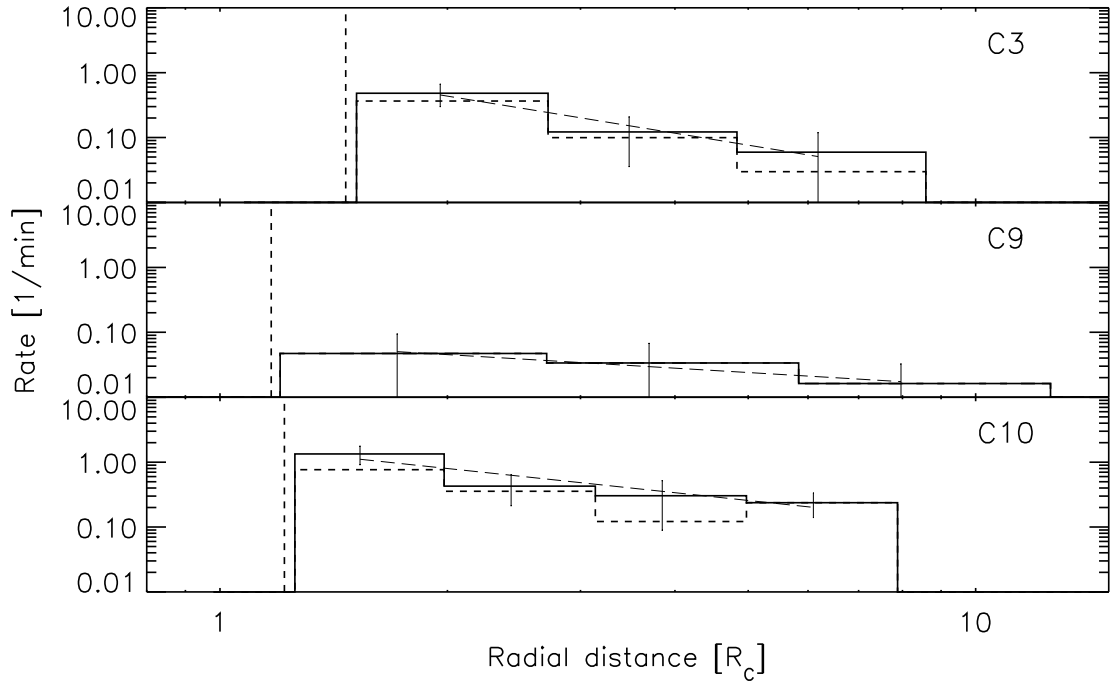}
        \caption{\label{rate_ca}
Impact rates of dust particles detected during three Callisto encounters.
Same notation as in Fig.~\ref{rate_eu}. Only particles with impact speed
$\rm < 10\,km\second^{-1}$ have been considered and no noise removal has been 
applied.
}
\end{figure}

With Callisto particles from all three Callisto encounters identified 
by their impact speed and impact direction, we can construct the 
radial profile of the dust impact rate in the same way as we did for 
the Europa flybys before. This is done in Fig.~\ref{rate_ca}. For the C3 
and C10 encounters the impact rate increased towards Callisto.
At C9 the number of detected particles was only three.
The derived radial profile --- although
being very uncertain --- is compatible with an increase towards Ganymede.
We conclude that
the dust impact rates detected at Callisto are 
compatible with a dust concentration surrounding this moon.

\bigskip
\centerline{\fbox{\bf Insert Figure \ref{rate_ca}}}
\medskip

\begin{figure}
\epsfxsize=0.8\hsize
\epsfbox{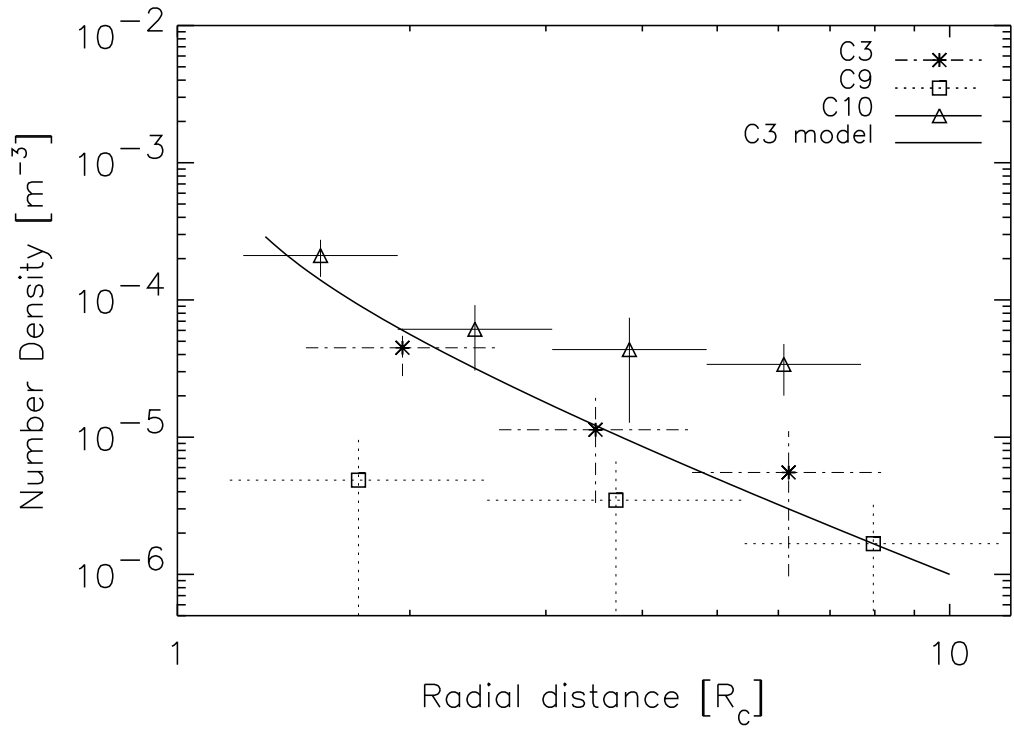}
        \caption{
\label{num_dens_ca}
Same as Fig.~\ref{num_dens_eu} but for Callisto. 
The theoretical profiles for C9 and C10 are very close to the one which is
plotted for C3.
}
\end{figure}

In Fig.~\ref{num_dens_ca} we show the number densities 
for Callisto derived from the radial profiles of the impact rate.
The data from the individual flybys show an increase towards
the surface of the moon. The number densities derived from 
the three flybys show a large variation from flyby to flyby, 
much larger than those for Europa. In particular, the number 
densities derived from the C9 data are very low. Possible 
reasons for this variation will be discussed in Sect.~\ref{sec_comp}.

In the same figure, the superimposed curves show the number density
profile calculated with the model \citep{krivov2003} and parameters
listed in Table~\ref{tab_parms}.
A comparison between the data and the model will be given in
Sect.~\ref{sec_comp}.

\bigskip
\centerline{\fbox{\bf Insert Figure \ref{num_dens_ca}}}
\medskip

\subsubsection{Mass distribution}

\begin{figure}
\epsfxsize=0.6\hsize
\epsfbox{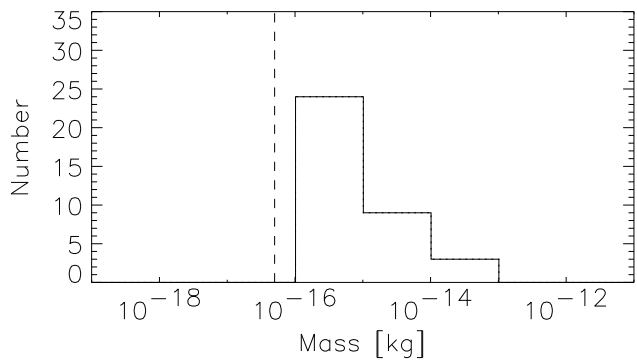}
\vspace{-8cm}

\epsfxsize=0.6\hsize
\epsfbox{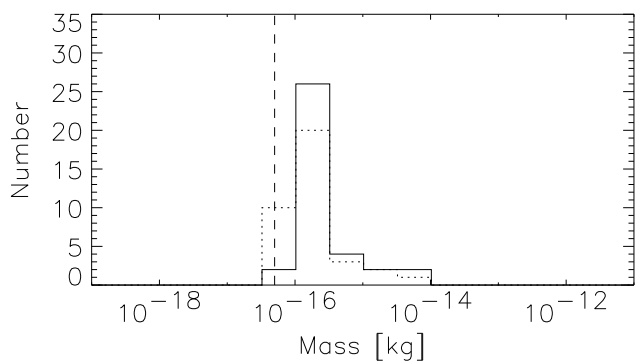}
        \caption{
\label{mass_hist_ca}
Same as Fig.~\ref{mass_hist_eu} but for three Callisto flybys.
The detection threshold is shown for $\rm 8 \km\second^{-1}$.
Only the 35 particles with a velocity error factor
$\rm VEF < 6$ \citep{gruen1995a} have been considered.
The dotted histogram is without aging correction, the data for 
the solid histogram
has been corrected for aging of the dust detector electronics.
}
\end{figure}

The mass distribution for the 35 Callisto particles with $\rm VEF < 6$ 
is shown in Fig.~\ref{mass_hist_ca}. As for Europa, we show the
mass distribution with the calibrated impact velocities (upper panel)
and that obtained by applying the velocity of Galileo relative 
to Callisto (bottom panel). Here the detection threshold is 
$\rm 8\,km\second^{-1}$, and the mass distribution is
incomplete around the threshold. Again, the mass distribution
is incomplete in the bins below about $\rm 10^{-15}\,kg$ due
to incomplete data transmission and the maximum of the mass 
distribution may be artificially too low by up to a factor of
1.3. 

\bigskip
\centerline{\fbox{\bf Insert Figure \ref{mass_hist_ca}}}
\medskip

The mean calibrated mass of the Callisto particles is 
$\rm 3.7 \times 10^{-16}\,kg$, which is an order of magnitude lower
than the value derived for the Europa cloud particles. Again, 
assuming spherical 
particles with a density of $\rm 1\,g\,cm^{-3}$, this 
corresponds to a particle radius of $ \rm 0.5 \,\mu m$. 
We have also corrected the masses of the particles for instrument
degradation and constructed a corrected 
mass distribution (solid histogram in the bottom panel of 
Fig.~\ref{mass_hist_ca}). 
The corrected mean mass of the Callisto particles is
$\rm 5.2 \times 10^{-16}\,kg$.

\subsection{Io}

\subsubsection{Impact direction}

\begin{figure}
\epsfxsize=0.9\hsize
\epsfbox{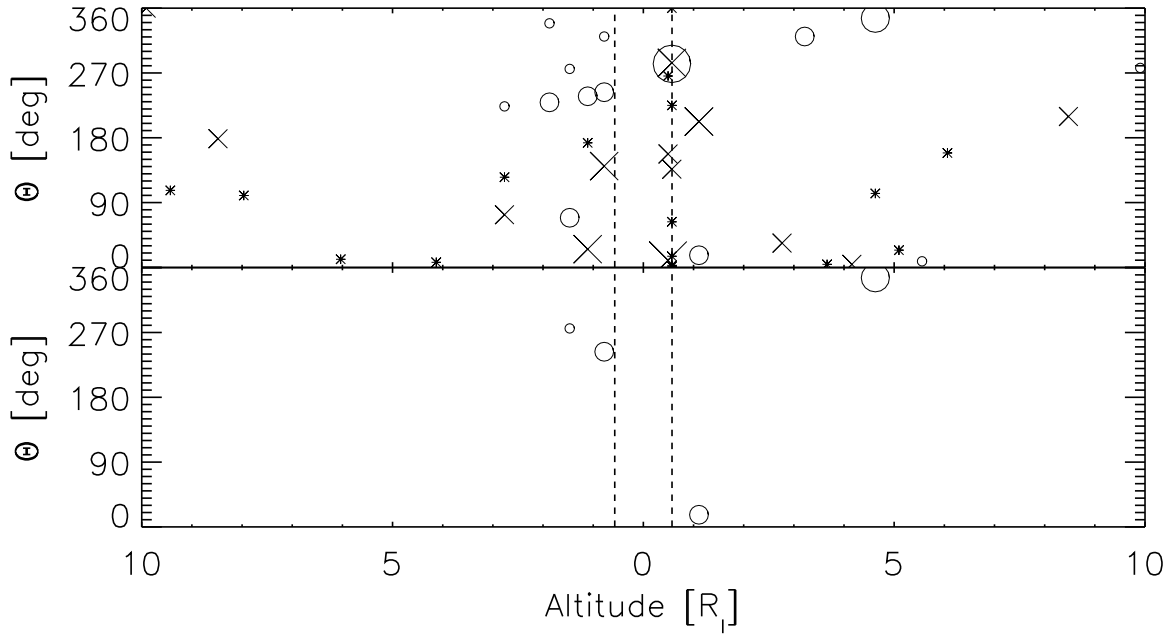}
        \caption{\label{rot_io}
Sensor direction (rotation angle, $\Theta$) versus altitude of the Galileo
spacecraft above the surface of Io at the time of dust impact for the I0 flyby.
The radius of Io, $\rm R_I$ is given in Table~\ref{phys_prop}.
The altitude range shown corresponds to a time interval of about 45~min.
Each symbol indicates an impact event (dust or noise), and
the size of the symbol indicates the generated  
charge ($ \rm 2 \times 10^{-14}\,C \leq Q_{\,I} \leq 10^{-8}\,C$).
Circles show class~2 and class~3, asterisks show class~1 in the lowest 
amplitude range, crosses denote class~1,
amplitude range 2 and higher. Top panel: all data in classes~1 to 3, bottom panel: 
class~3 and noise-removed class~2 data. 
}
\end{figure}

Galileo had a total of seven flybys at Io but only the initial one
in December 1995 (I0; the orbit notation is I ''zero``) 
had a favourable detection geometry. During 
the other flybys at this satellite which occurred after mid-1999 
the sensor orientation prevented the 
detection of ejecta dust particles. Figure~\ref{rot_io} shows the
sensor orientation at particle impact at the I0 encounter. The bottom
panel shows class~3 and noise-removed class~2 data (only four impacts).

The noise identification criteria applied to the Europa data 
have been developed for the spatial region outside $\rm 10\,R_J$. 
The I0 data, however, have been collected closer 
to Jupiter at $\rm 6\,R_J$ where the noise characteristics may have 
been different \citep{krueger1999c}. We therefore 
show the full data set of classes~1 to 3 in the top panel of 
Fig.~\ref{rot_io}. The noise fraction in class~2 derived with the 
algorithm for secondary ejecta grains is about 80\%. The
complete class~2  data set also shows a concentration of 
grains towards Io. Therefore, the noise rejection algorithm 
may be too restrictive, thus  rejecting too many 
events. Class~1 events, which are usually classified as 
noise in the Jovian environment, show an interesting behaviour:
the impacts cluster at rotation angles 
$\rm 0 \leq \Theta \leq 180^{\circ}$. This direction is
compatible with the approach direction of plasma particles in the 
Io torus. 

A few days before the flyby at Io, the channeltron 
high voltage was decreased and the charge detection thresholds
were raised \citep{gruen1996c,krueger1999a} to reduce the instrument 
sensitivity in the high radiation environment of the inner Jovian 
magnetosphere. This reduced the instrument sensitivity for class~3 
impacts. Those impacts, however, that did not generate enough 
charge to become class~3 events should have shown up as class~2 
impacts. 
Unfortunately, class~2 is contaminated with noise so that the 
identification of these particles is ambiguous. Fortunately, 
only a small fraction of the data sets of particles was lost
due to incomplete data transmission (Table~\ref{tab_sat}).

\bigskip
\centerline{\fbox{\bf Insert Figure \ref{rot_io}}}
\medskip
%*MS* Also obsolete reference to Table II
%*MS* \bigskip
%*MS* \centerline{\fbox{\bf Insert Table \ref{tab_sat}}}
%*MS* \medskip

Figure~\ref{rot_io} shows four particles within an altitude 
of about $\rm 6\,R_I$ (the Hill sphere of Io). Two of these 
particles approached the detector from a direction $\Theta \approx
270^{\circ}$. The other two particles were detected when the 
dust detector pointed $\approx 90^{\circ}$ away from this direction. 
A check of the approach direction of potential Io particles revealed
that all four particles are compatible with an Io origin: the
approach direction of Io particles as seen from Galileo was so 
close to the spin axis of the spacecraft (direction opposite to 
antenna axis) that they were detectable at almost all rotation 
angles \citep{krueger1999c}. We therefore consider all four particles as 
probable Io particles. Only one of these impacts occurred in
ion amplitude range AR1, two were detected in AR2 and one in AR3. 
Io dust stream particles were detected in AR1 only \citep{gruen1998}
so that the identified particles were unlikely stream particles.
In addition, the impact rate of stream particles was reduced in 
the inner Jovian system because of reduced impact speeds 
\citep{graps2001a}.

\subsubsection{Impact velocity}

The total number of identified potential ejecta particles from
Io is only four. With this very low number of detections no reasonable
velocity distribution can be constructed. We can, however, still
check whether the average particle velocity is compatible
with the hypothsis that the particles are basically at rest with 
respect to Io. The
velocity of Galileo relative to Io at the I0 encounter was 
$\rm 15\,km\second^{-1}$. The averaged particle speed derived for
the four particles is $\rm 10.3 \pm 8.4\,km\second^{-1}$. Although
one has to keep in mind that the statistical uncertainty of this
value is very large, it is compatible with particles being 
bound to Io.

\subsubsection{Impact rate and number density}

\begin{figure}
\epsfxsize=0.9\hsize
\epsfbox{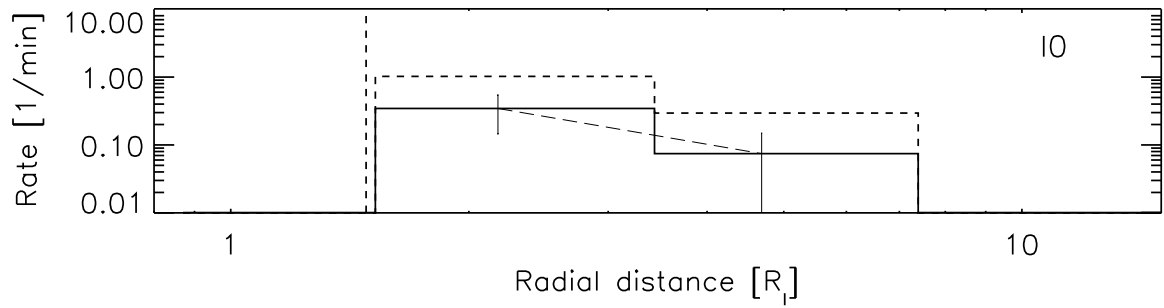}
        \caption{\label{rate_io}
Impact rates of dust particles detected during Galileo's I0 encounter of Io.
Same notation as in Fig.~\ref{rate_eu}. Solid histogram: class~3 and noise-removed class~2 
data (derived from only four dust impacts), dashed histogram: class~3 and all class~2 data 
(without noise removal).
}
\end{figure}

The dust impact rate at Io derived from the four identified Io
particles is shown in Fig.~\ref{rate_io}. It shows a slight
concentration towards Io. One has to keep in mind, however,
that the radial profile of the impact rate is very uncertain
because of the small number of dust detections in the close 
vicinity of this moon.

\bigskip
\centerline{\fbox{\bf Insert Figure \ref{rate_io}}}
\medskip

\begin{figure}
\epsfxsize=0.9\hsize
\epsfbox{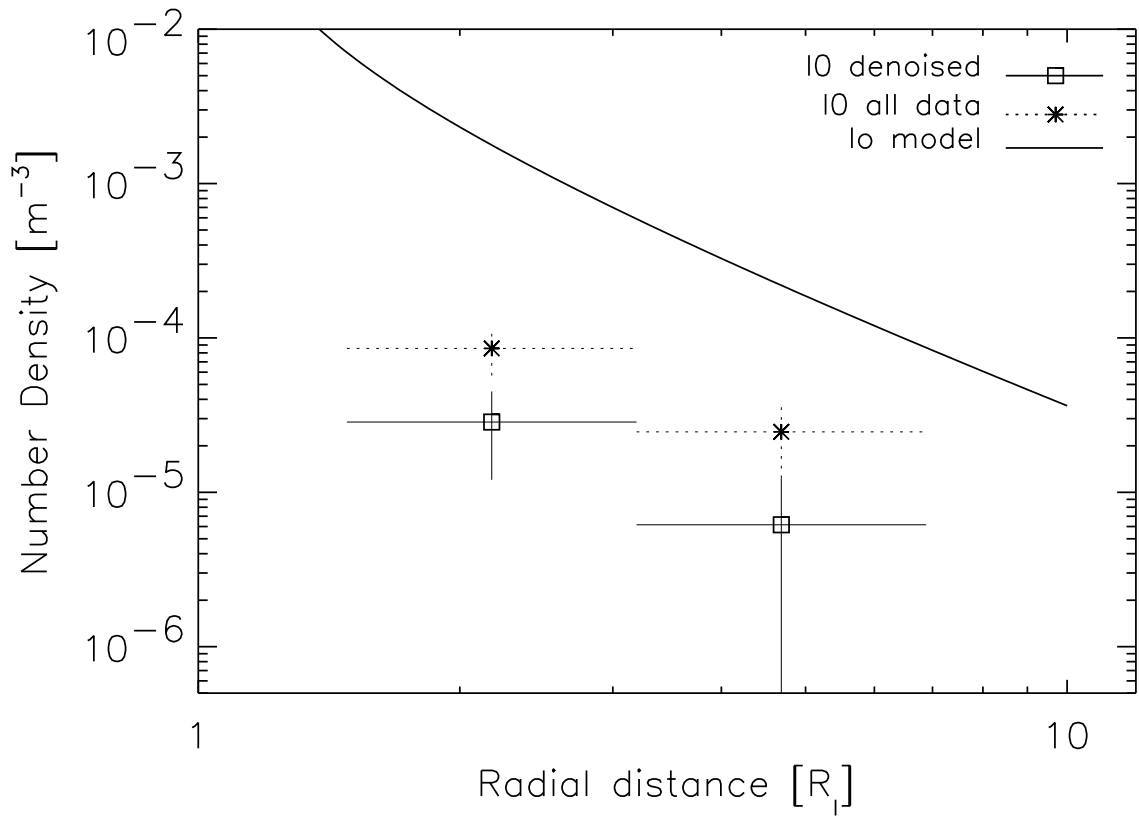}
        \caption{
\label{num_dens_io}
Same as Fig.~\ref{num_dens_eu} but for Io. Class~3 and noise-removed class~2 data
are shown with solid lines. All class~3 and class~2 data taken together
are shown with dotted lines.
}
\end{figure}

\bigskip
\centerline{\fbox{\bf Insert Figure \ref{num_dens_io}}}
\medskip

The number density of dust in the close 
vicinity of Io derived from the impact rate profiles
is shown in Fig.~\ref{num_dens_io}. For comparison we
show the data points for the noise-removed data set (solid
lines) and for the complete class~2 and class~3 data 
set. 
The curves in the same figure depict the number density
profile calculated with the model \citep{krivov2003} and parameters
listed in Table~\ref{tab_parms}.
A comparison between the data and the model will be given in
Sect.~\ref{sec_comp}.

\subsubsection{Mass distribution}

No attempt has been made to construct a mass distribution because
of the small number of detections at this moon. The mean
mass of the particles taking the speed of Galileo relative
to Io as the impact speed 
is $\rm 8.5 \times 10^{-16}\,kg$. Note that this value
is not affected by the instrument degradation because this Io
flyby occurred at the beginning of Galileo's Jupiter mission.

\section{Comparison of the dust clouds surrounding the 
Galilean moons}   \label{sec_comp}

In the previous section we have analyzed the 
dust impacts detected in the circumsatellite dust clouds
individually for Io, Europa and Callisto. The dust 
cloud of Ganymede has been investigated in an 
earlier publication \citep{krueger2000a}.
For each of the four moons we have
identified impacts of probable ejecta cloud 
particles in the Galileo dust data set, determined their
impact speeds and mass distributions and have derived 
impact rate and number density profiles.
We now take the data sets for all four moons together to
compare the properties of their surrounding dust clouds.

\subsection{Mass distribution}

\begin{figure}
\epsfxsize=0.7\hsize
\epsfbox{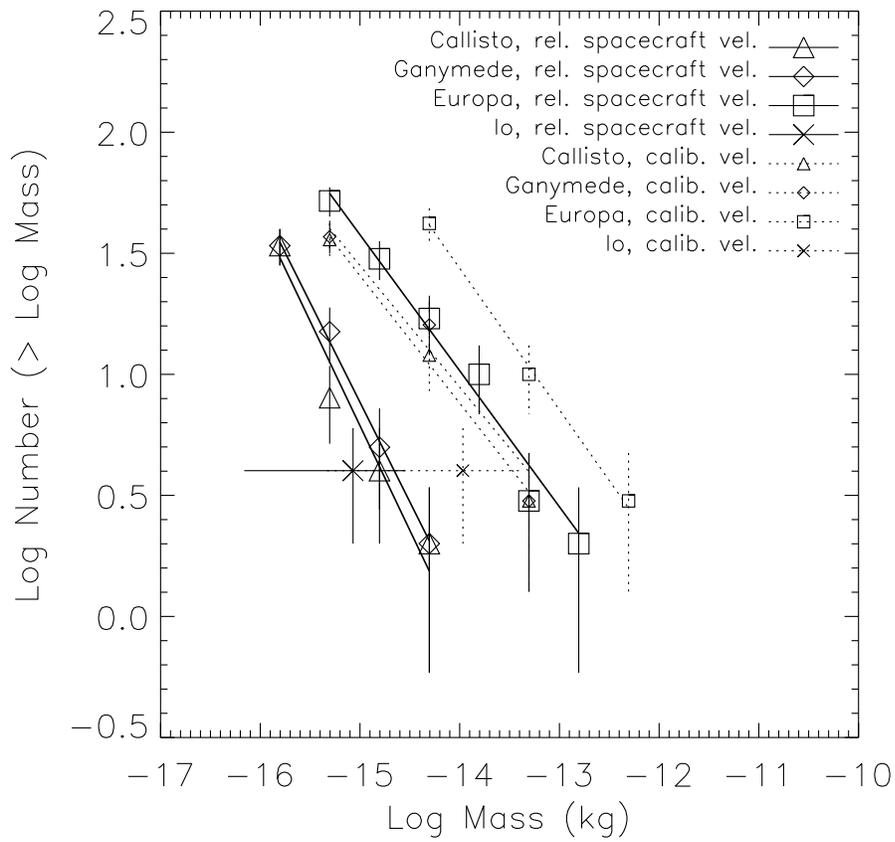}
        \caption{
\label{massdist}
Cumulative mass distributions from Fig.~\ref{mass_hist_eu} and \ref{mass_hist_ca} 
for Europa and Callisto, respectively, and from \citet{krueger2000a} for
Ganymede. 
For Io only one data point is shown because of the sparcity of the data.
Straight lines are linear fits to the data. 
Vertical bars indicate the $\sqrt{N}$ statistical uncertainty. 
}
\end{figure}

The mass distribution of the grains allows a simple check 
for the compatibility of the data with the hypothesis
of the impact origin of the detected particles.
We took the mass distributions for Europa, Callisto and Ganymede
(Fig.~\ref{mass_hist_eu}, Fig.~\ref{mass_hist_ca} and 
\citep{krueger2000a}) and show linear fits to the cumulative 
distributions in Fig.~\ref{massdist}.
%We only considered grains with masses 
%well above the detection threshold ($M > 10^{-16}\kg$).
The slopes of the cumulative mass distributions assuming
Galileo's velocity relative to each moon as 
the particle impact speed are given in Table~\ref{tab_results}.
Only one data point is shown for Io because of the scarcity of the data.
For the three other moons the slopes are in the range between 
0.58 and 0.86, which is in good agreement with the typical slopes 
one expects for impact ejecta ($0.5 \la \alpha \la 1.0$;
see, \eg, \citet{koschny2001b}). It should be emphasized that
even though the statistical uncertainties in the data sets are
relatively large because of the small number of detections, the
slopes derived for the three moons do not differ very much. 
The slopes derived for the mass distributions taking the
calibrated impact speeds are in the range 0.5 -- 0.6 and 
are thus even closer together, although they are flatter than 
those derived with the spacecraft speed relative to the moon.

It should be noted that the mass distributions of dust around Ganymede and
Callisto agree very well (0.82 vs. 0.86) whereas the one for 
Europa is somewhat flatter (0.58).
This might reflect differences in surface properties of the satellites:
for instance, flatter mass distributions are typical of looser targets than
of consolidated ones 
\citep[see, \eg,][and references therein]{koschny2001b}.

\bigskip
\centerline{\fbox{\bf Insert Figure \ref{massdist}}}
\medskip

\bigskip
\centerline{\fbox{\bf Insert Table \ref{tab_results}}}
\medskip

\subsection{Number densities}
\label{subsec_numden}

\begin{figure}%[tbh]
\epsfxsize=0.7\hsize
\epsfbox{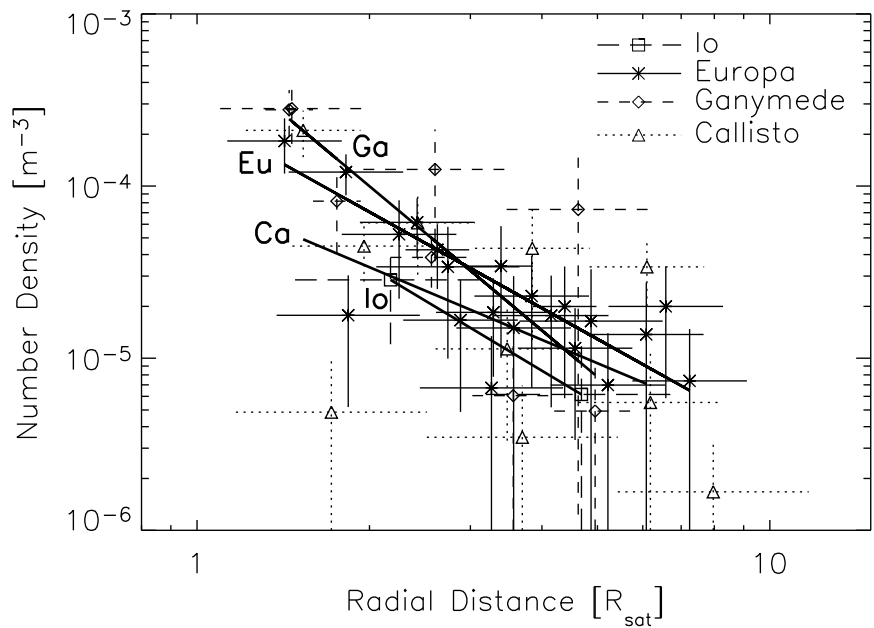}
\caption{
\label{num_dens_all}
  Number density of dust as a function of radial distance from the center
  of Ganymede (data from 4 flybys), Europa  (8 flybys),
  Callisto (3 flybys) and Io (1 flyby). The altitude is shown in units of
  the satellite radius (Tab~\ref{phys_prop}).
  Vertical error bars reflect statistical uncertainty due to the small
  number of impacts.  The straight lines are mean measured curves for
  each moon (Table~\ref{tab_results}, col.~8) 
}
\end{figure}

The number densities derived for all four Galilean moons 
are shown in Fig.~\ref{num_dens_all}.
Straight lines are least squares fits to the data for each moon.
We list the slopes of these curves, which are averages of slopes
for individual flybys at each moon, in Table~\ref{tab_results} (col.~8).
The average slopes for each moon
are between $-1.6$ for Callisto and $-2.8$ for Ganymede, 
with Europa being in between ($-2.0$).
The very uncertain slope for the Io data ($-2.0$) is close to the 
Europa value.
The Ganymede data show the steepest slope but also the largest
uncertainty.
This is mainly caused by the incomplete data 
transmission which mostly affected the G2 and G8 measurements.
The incompleteness affects the error via 
$\propto(N\pm\sqrt{N})\times{\rm correction}$, therefore giving
larger limits compared to the $100\%$ transmission case.

Altogether, the slopes are roughly consistent with the one
predicted in the framework of the spherically-symmetric cloud
%*MS* Again power law slopes... Instead of -2 up to -2.5
%*MS* it could read -2 up to -3. For r<5 R_sat the number density is
%*MS* really steeper than r^(-2.5). But I did not change the
%*MS* text here.
model \citep{krivov2003}, between $-2.0$ and $-2.5$ 
[see Eqs.~(\ref{bound_cal})--(\ref{unbound_cal})].
Only the slope derived for Callisto is somewhat flatter. 
A study of possible asymmetry effects in the clouds
has shown that this ``reference'' slope may be substantially flatter
or steeper, depending on the flyby geometry and the position of
the satellite in its orbit at the time of flyby \citep{sremcevic2003}.
This may account for a tangible scatter in the slopes that we derived
from the data.

The measured absolute number densities of all four clouds at a given distance
(measured in satellite radii) are similar~---
see Fig.~\ref{num_dens_all}. It is important to compare this result
with the theory.
The absolute number density of a dust cloud at a given distance
from a satellite should depend in a non-trivial way on a number of factors:
mass of the moon, its planetocentric distance (through a
distance-dependent gravitational focussing of the impactor flux),
as well as the satellite surface properties. All these dependences are
taken into account in the model which was confined, however, to
a solid ice-silicate surface \citep{krivov2003}.
%-AVK My attempt to respond to Josh's comment 14!!!
%As a result, the number densities of the clouds around Europa, Ganymede and
%Callisto, computed with the model, turned out to be close to each other and
%to be consistent with the densities derived from the data.
The number densities of the clouds around Europa, Ganymede and
Callisto, computed with the model, turned out to be within a factor of
several from each other (most notably, larger yields $Y$ for closer-in satellites
are compensated by lower ejecta speeds $u_0$, because of an energy conservation
requirement used in the model).
The number densities of these three clouds
are consistent with the densities derived from the data.
Not so for the Io cloud: the same model
\citep{krivov2003} predicts a much higher
dust number density (mostly because of the larger flyby speed at Io, resulting
in a possibility of detecting much smaller dust grains compared to the other
Galileans).
The number density in the Io cloud predicted by the model is
at least an order of magnitude higher than observed. This could be due to
the scarcity of the Io data (4 individual dust impacts only).
Alternatively, this may be a real effect,
caused by different surface properties of Io (volcanic deposits and
condensed gases like $SO_2$ frost)
compared to the other three Galilean moons (presumably ``solid'' ice
with some contamination by non-volatile materials).
Such a view seems to be indirectly supported by our preliminary
analysis of the dust environment between the orbits of Io and Europa
(work in progress):
the density of the ``Galilean ring'' \citep{krivov2002a} does not seem to
increase from the Europa orbit towards Jupiter, which might be compatible
with Io being a weaker source of ejecta than it would be, if it were
similar in surface properties to the other Galilean satellites.

\bigskip
\centerline{\fbox{\bf Insert Figure \ref{num_dens_all}}}
\medskip
  
A comparison of data from different flybys at the same moon 
shows that for Europa the derived dust densities do not vary 
significantly from flyby to flyby. This indicates little or 
no temporal variation and/or dust density variation between the 
leading and the trailing side of this moon with respect to the field
of impactors. On the other hand, the three Callisto flybys showed a
significant variation by  more than an order of magnitude between the C9
and the C10 flybys.  For Ganymede, the G2 and G8
flybys showed  somewhat  larger number densities than the G1 and G7
flybys. The former, however, have the largest uncertainties because
of incomplete data transmission and particle identification via
the impact speed criterion.
At present, it is not clear whether the differences between the data
from different flybys of Callisto and Ganymede can be attributed to
asymmetries in the circumsatellite dust clouds modelled 
in \citet{sremcevic2003}. A comparison between the data and theory is
hampered by poorly known directionality of impactors in the vicinity of
Jupiter. This issue will be the subject of a future investigation.

\subsection{Grain velocities in the clouds}

The theoretical models of dust clouds predict that, on average,
the constituent particles should have substantial velocities with respect
to their parent satellites. Based on \citet{krivov2003} formulas, about
40--50\% of the grains at distances from 2 to 8~$R_{\mathrm{sat}}$
in the Ganymede and Callisto clouds should be faster that
$2\km\second^{-1}$. For Europa the fraction is 30--40\%.

The question that we address now is: can we find indications in the
data that some grains have appreciable velocities relative to the
respective moon? The most natural way would be to look at possible
deviations of the actual impact speeds from the mean value, equal to the
spacecraft speed with respect to the satellite. Unfortunately, this
is not possible: as we have seen (Figs.~\ref{velocities_eu} and
\ref{velocities_ca}), the instrument calibration is by far not
accurate enough to do that. Another possibility would be to look at
the impact directions, \ie\ at the rotation angles of impacts.
For most of the flybys the cloud particles, 
%HK: removed: assumed to ``levitate'' in the cloud
%-AVK ... next line added to replace a "levitating phrase":
if they were at rest with respect to the moon,
could only be detected in the rotation angle range
$\Theta = 270^{\circ} \pm x$, where the semi-width $x$ of the
detectability range is a (known) function of the FOV opening angle
and the angle between the Galileo ram direction and its spin axis,
$\beta$.
The semi-width $x$ is $180^\circ$ for $\beta \le 10^\circ$ and decreases
to $67^\circ$ at $\beta = 90^\circ$.
Should a particle, which we identify as a cloud particle, have had
a $\Theta$ value somewhat outside the range
$\Theta = 270^{\circ} \pm x$, this would be an indication
that the particle had an appreciable velocity, so that the impact velocity
deviated markedly from the anti-ram direction.

One should not expect the number of such events to be high: even very fast
grains can only show up in this test if the direction of their velocity
vector is appropriate, and only for some flyby geometries.
%HK: changed: 
We have checked all cloud particles and found four individual
impacts of this kind: one in G7, one in E11 and two in E12.
%-AVK The problem is that, counterintuitively, one should expect a
%-AVK larger relative velocity... farther out from the moon! I strongly
%-AVK suggest to remove the complete sentence to avoid confusion!
%Interestingly, the
%two impacts in E12 were close to the surface where one expects the
%largest relative velocity with respect to the satellite.
%HK: end changed text

\subsection{Mass budget}

As was the case for Ganymede earlier \citep{krueger2000a}, 
we give some general estimates concerning the
mass budget of the dust clouds of the other Galilean moons as derived 
from the model \citep{krivov2003}.
The results, including new estimates for Ganymede,
are collected in Table~\ref{tab_budget}.
Note that these are only crude estimates which are uncertain by at least one
order of magnitude, perhaps even more.
The expected steady-state masses of the clouds range from about 10 tons
for Callisto to about 200 tons for Io; the Io cloud may, however, be
lighter~--- see discussion in Sect.~\ref{subsec_numden}.
Interestingly, the mass injection rate of the material into circumjovian
space is similar for all four moons and is, in turn, comparable to
the mass flux of impactors onto respective satellites, $\sim
100\g\second^{-1}$. 
This means that each satellite ``redirects'' nearly as
much dust into the circumplanetary space as it receives from the
interplanetary one. Of course, the mass/size and velocity distributions of
the ``incoming'' and ``outgoing'' matter are generally quite different.

\bigskip
\centerline{\fbox{\bf Insert Table \ref{tab_budget}}}
\medskip

%=========================================================================

\section{Conclusions}
\label{sec_conclusions}

We have examined the dust impacts registered by the Galileo
dust detector in the immediate vicinity of Io, Europa and Callisto
during a total of 12 flybys at these Jovian moons. 
By analyzing impact directions and velocities and the mass distribution, 
as well as spatial locations of the dust impacts in 
comparison with model predictions \citep{krivov2003}, 
we have shown that the particles originated from the moons. 
Our analysis technique was similar to an earlier investigation 
of dust data collected at Ganymede \citep{krueger2000a}. The
dust impacts recorded at all four moons are compatible with impact 
debris produced by hypervelocity impacts onto the surfaces of these
moons. 
%The projectiles responsible for the formation of the 
%circumsatellite clouds of dust grains are most likely 
%interplanetary micrometeoroids. 
For the icy moons 
Europa, Ganymede and Callisto, the mass distributions of the detected 
grains, as well as the spatial dust densities derived from the
measurements are in fairly good agreement with 
the predictions from the model of hypervelocity impacts of
interplanetary dust particles (IDPs),
assuming contemporary models of IDP flux at a heliocentric distance
of Jupiter and a low-temperature ice-silicate target. For Io, the 
number of dust detections is too small to derive a reliable 
mass distribution. The number density obtained for this moon
is more than an order of magnitude lower than predicted by the
model which assumes a solid ice surface. The lack of detections
may be due to a softer, fluffier surface of Io
compared to the three icy Galilean moons. Io's surface 
is (at least partially) covered with volcanic deposits.

This work continues the analysis of the dust clouds 
surrounding the Galilean moons and confirms the previous scenario 
of ejecta dust clouds generated by hypervelocity impacts of 
micrometeoroids. Up to now, this had only been tested at Ganymede.
Our theoretical description is based on
the physical conditions in the Jovian system as well as available
laboratory data of hypervelocity impacts. We have neglected any
spherical asymmetries of the dust clouds surrounding the moons. 
To a first approximation, this is supported by the data, 
especially the measurements taken at Europa. On the other hand,
the Callisto and possibly Ganymede measurements show a variation between
different flybys which might be indicative of a leading-trailing cloud
asymmetry caused by the motion of the moons through the field
of impactors, as predicted by theory \citep{sremcevic2003}.
This will be addressed in a future analysis.

Most of the dust ejected from the surface is launched into
bound orbits and falls back to the moon. These 
short-lived, but continuously replenished grains form the 
ejecta dust clouds. A tiny fraction of impact debris is ejected 
at speeds sufficient to escape from the moon entirely. 
The ejected mass is comparable with the incoming flux of IDP 
impactors. The escaping grains go into orbit about Jupiter
and most of them will eventually be swept up by one of the Galilean 
satellites. A tiny fraction of them forms a tenuous dust ring
surrounding the planet \citep{krivov2002a}. This ring is by 
far too tenuous to be detected optically. By the impact ejecta 
mechanism, moons turn out to be efficient sources for dusty
planetary rings. 
In particular, Jupiter's gossamer ring 
and Saturn's E ring are thought to be maintained by ejecta particles
from smaller moons which orbit their parent planets within the rings. 
In November 2002 Galileo traversed the gossamer
ring and had a close flyby of Amalthea, one of the small Jovian 
moons which orbits the planet within the ring region. The dust measurements
collected during this passage may give new insights into the dynamics and 
feeding mechanism of this dusty ring and about the significance of
small moons as sources of dust.

All celestial bodies without gaseous atmospheres (asteroids and planetary 
satellites of all sizes) should be surrounded by an ejecta dust cloud. 
The dust particles in the cloud are composed of surface material from
the parent body and, hence, carry information about the surface from 
which they have been kicked up. Our analysis of the Galileo in-situ dust 
data has shown that spacecraft measurements near celestial bodies ---
which act as sources of dust --- can be used as a new diagnostic tool 
to analyze the surface properties of these bodies. This is of particular 
interest for the Cassini mission which will investigate the Saturnian 
system beginning in 2004.
The Cassini dust instrument will
be able to measure the chemical composition of particles in the
dust clouds surrounding the Saturnian moons. This way,
the surface properties of the source moons can be investigated 
remotely. Interestingly,
the in-situ dust measurements turn into a remote 
sensing technique where the dust instrument is used like a telescope 
for surface investigation.

\bigskip
{\bf Acknowledgements.}
The authors wish to thank Frank Spahn for many valuable discussions.
This research has been supported by the German Bundesministerium f\"ur Bildung 
und Forschung through Deutsches Zentrum f\"ur Luft- und Raumfahrt e. V. 
(DLR, grants 50~QJ~9503~3 and 50~OH~0003)
and by the Deutsche Forschungsgemeinschaft (DFG, grant No. Sp~384/12-3).
We wish to thank the Galileo 
project at JPL for effective and successful 
mission operations.

\newpage
%=========================================================================

%\bibliography{/home/krueger/tex/bib/pape,/home/krueger/tex/bib/references}

%=========================================================================

\newpage

%\section*{Figures}

%---------------------------------------------------------------------------

%---------------------------------------------------------------------------

%---------------------------------------------------------------------------

%---------------------------------------------------------------------------^M

%---------------------------------------------------------------------------^M

%---------------------------------------------------------------------------

%---------------------------------------------------------------------------

%---------------------------------------------------------------------------^M

%---------------------------------------------------------------------------^M

%---------------------------------------------------------------------------^M

%---------------------------------------------------------------------------

%---------------------------------------------------------------------------^M

%---------------------------------------------------------------------------^M

%---------------------------------------------------------------------------

%---------------------------------------------------------------------------

%---------------------------------------------------------------------------

\clearpage

\begin{table}[tbh]
\caption{
Physical properties of Jupiter and the Galilean satellites used in this paper.
The Hill radius is defined as $r_{\rm Hill} = r (\frac{m}{3(M + m)})^{1/3}$
with $M, m$ being the masses of Jupiter and the moon, separated by distance 
$r$.
}
\label{phys_prop}
\begin{tabular}{lccccc}
\\
\hline
\hline
Object & Jovicentric &   Radius         & Symbol  & Hill radius     & Escape speed     \\
       & distance    &  $r_{\rm obj}$   &         & $ r_{\rm Hill}$ &$v_{\rm esc}$ \\
       & ($\rm R_J$) &     (km)         &         & ($\rm R_{obj}$) &($\rm km\,s^{-1}$)\\
\hline
Jupiter & --         &     71,492       &$\rm R_J$&     --          &      --          \\
Io      &\,\,\,5.9   &\,\,\,1,818       &$\rm R_I$&\,\,\,5.8        &     2.56         \\
Europa  &\,\,\,9.4   &\,\,\,1,560       &$\rm R_E$&\,\,\,8.7        &     2.03         \\
Ganymede&     15.0   &\,\,\,2,634       &$\rm R_G$&   12.0          &     2.74         \\
Callisto&     26.3   &\,\,\,2,409       &$\rm R_C$&   20.9          &     2.44         \\
\hline
\hline
\end{tabular}
\end{table}

%HK: I have added the slopes for E6, E14,E15 and E17 in this table.
\begin{table}[tb]
\caption{
Galileo flyby characteristics and parameters for the dust particles detected 
%within $\rm 8\,R_E$ altitude 
during Galileo's flybys at Europa, Callisto and Io
(for Europa, particles within $\rm 8\,R_E$ altitude have been included, and 
for Callisto and Io within $\rm 6\,R_C$ and $\rm 6\,R_I$, respectively): 
Flyby number (col. 1),
time of flyby (col. 2),
altitude at closest approach to satellite (col. 3), 
velocity of Galileo relative to satellite (col. 4), 
average measured particle velocity (velocity error factor $\rm VEF < 6$,
\citet{gruen1995a}; col. 5),
spin-averaged sensor area (maximum value $\rm 235\,cm^{2}$; col. 6),
number of class~2 and class~3 satellite particles for which their complete data set has 
been transmitted to Earth ($\rm 180^{\circ} < \Theta < 360^{\circ}$; col. 7),
number of all events (dust plus noise) detected with $\rm 180^{\circ} < \Theta < 360^{\circ}$ 
(both within $\rm 8\,R_E$, col. 8),
completeness of data set due to incomplete data transmission of Galileo (col. 9) and
slope of power law fit to the radial variation of the impact rate
(Fig.~\ref{rate_eu}, \ref{rate_ca}, \ref{rate_io}; col. 10).
The slopes are weighted with the square root of the number of 
particles, and the uncertainty takes into account the error bar of each data point.
Values given in parentheses have been derived from only four or fewer particles.
}
\label{tab_sat}
\footnotesize
\hspace{-0.4cm}
\begin{tabular}{cccccccccc}
\\
\hline
\hline 
Flyby& Date     &Altitude         & Spacecraft & Average       & Sensor     & Particles  & All events     & Complete- & Slope of \\
     &          &                 & velocity   & particle      & area       & with full  & with full      & ness of   & impact \\
     &          &                 &            & velocity      &            & data sets  & data sets      & data set  & rate   \\
     &(Year-Day)&   (km)          &($\km\second^{-1}$)& 
                                      ($\km\second^{-1}$)   & ($\rm cm^2$)&        &    &     (\%)  &        \\
  (1)&   (2)    &   (3)           &   (4)      &   (5)      &  (6)          &  (7)       & (8)       &   (9)      &  (10)   \\
\hline
     &          &         &           &               &        &                &                  &                &                \\[-1.0ex]
   \multicolumn{10}{c}{Europa} \\[0.5ex]
 E4  &96-354.287&\,\,\,\,\,698&5.7   & $6.8 \pm 4.0$ & 233   &      18        &      23    &    86     & $-2.21\pm 0.36$\\
 E6  &97-051.713&\,\,\,\,\,586&5.7   & $5.9 \pm 4.0$ & 224   &\,\,\,3         &\,\,\,7     &    88     & ($-1.69\pm 0.58$)\\
 E11 &97-310.855&  2,043  &    5.6   & $5.5 \pm 2.2$ & 235   &      10        &      11    &    90     & $-1.43\pm 2.25$\\
 E12 &97-350.502&\,\,\,\,\,201&6.3   & $4.5 \pm 1.7$ & 142   &      12        &      13    &    61     & $-2.72\pm 0.91$\\
 E14 &98-088.556&  1,644  &    6.5& $3.4 \pm 1.3$ &\,\,\,\,61 &\,\,\,3         &\,\,\,4    &   100     & ($-1.26\pm 0.44$)\\
 E15 &98-151.884&  2,515  &    6.4   & $5.5 \pm 1.5$ & 155   &\,\,\,4         &\,\,\,6    &   100     &  ($-2.11\pm 0.59$)\\
 E17 &98-269.163&  3,582  &    6.0   & $2.8 \pm 0.8$ & 180   &\,\,\,4         &\,\,\,4    &    54     &  ($+0.27\pm 1.56$)\\
 E19 &99-032.097&  1,439  &    5.8   & $5.7 \pm 4.7$ & 152   &\,\,$10^{\dagger}$&\,\,$13^{\dagger}$&$100^{\dagger}$ & $-1.70\pm 0.62$\\
     &          &         &          &               &        &                &                  &                &                \\[-0.5ex]
   \multicolumn{10}{c}{Callisto} \\[0.5ex]
 C3  &96-309.566&  1,118  &    8.0   & $6.3 \pm 1.7$ & 224    &   10         &  35        &   80      &$-1.90 \pm 0.45$\\
 C9  &97-176.575&\,\,\,\,\,415&8.2   & $7.6 \pm 0.2$ & 197    &\,\,3         &\,\,5       &   100     &($-0.69 \pm 0.10$)\\
 C10 &97-260.013&\,\,\,\,\,538&8.2   & $6.3 \pm 2.3$ & 142    &   22         &  37        &   64      &$-1.31 \pm 1.06$\\
     &          &         &          &               &        &              &            &           &                \\[-1.0ex]
   \multicolumn{10}{c}{Io} \\[0.5ex]
 I0  &95-341.740& 892     &  15.0      &$10.3 \pm 8.4$ & 133  &    4         &   13       &   87      &($-2.00 \pm 1.88$)\\
\hline
\hline
\end{tabular}
\normalsize
\mbox{}\\
\vspace{2mm}
$\dagger$: Data transmission ceased 1 Feb 99, 02:38:46h, \ie\ 19~min after
closest approach, 
at 6,767\,km altitude. All data sets of particles detected earlier during the E19 encounter 
were transmitted.
\end{table}

\begin{table}[tbh]
\caption
{
Model parameters for different satellites:
speed of impactors,
geometric albedo,
assumed silicate content,
energy partitioning parameter,
characteristic yield,
parameters of the ejecta speed distribution
(see Krivov et al. 2002b for description of the parameters).
}
\bigskip  
\begin{center}
\begin{tabular}{lccccccc}
\hline  
\hline  
Satellite                     &
$v_{\rm imp}$                 &
$A$                           &
$G_{\rm sil}$                 &
$K_{\rm e}/K_{\rm i}$         &
$Y$                           &
$u_0$                         &
$\gamma$                      \\
                      &
$(\km\second^{-1})$    &
                      &
                      &
(\%)                  &
                      &
$(\m\second^{-1})$     &
                      \\
\hline
Io       & 26 & 0.61 &  0 & 30 & $2.8\times 10^4$ & 28 & 2.0\\
Europa   & 21 & 0.64 &  0 & 30 & $1.6\times 10^4$ & 30 & 2.0\\
Callisto & 15 & 0.20 & 70 & 20 & $7.1\times 10^2$ & 51 & 1.4\\
\hline  
\hline
\end{tabular}
\end{center}
\label{tab_parms}
\end{table}

\begin{table}[tbh]
\caption{
Derived properties of the impact-generated dust clouds (data for Ganymede 
taken from \citet{krueger2000a}): mean flyby speed of Galileo at the
satellite (col.~2), number of detected cloud particles for which the complete data
set has been transmitted (col.~3), mean measured impact speed of these 
particles ($\rm VEF < 6$; col.~4), mean particle mass (col.~5), corresponding 
particle radius assuming spherical particles with density $\rm 1\,g\,cm^{-3}$ 
(col.~6), slope of the power law fit to the cumulative mass distribution (col.~7) 
and average slope of radial number density distribution (col.~8; averages of 
slopes from individual flybys).
The speed of Galileo relative to the satellite has been assumed for the
values in col.~5 to 7. For the uncertainties in the number density slopes (col.~8)
the uncertainties of the slopes from each individual flyby have been taken into 
account.
}
\label{tab_results}
\small
\begin{tabular}{lccccccc}
\\
\hline
\hline
Object & Average            & Number &  Average        & Average                & Average      & Slope $\alpha$ & Slope of\\ 
       & flyby              & of     &  particle       & particle               & particle     & of mass        & radial number\\
       & speed              &detections&  speed        & mass                   & radius       & distrib.       & density distrib.\\
       &($\rm km\second^{-2}$)& &($\rm km\second^{-2}$)& (kg)                   &($\rm \mu m$) &                &         \\
\multicolumn{1}{c}{(1)}& (2)&    (3) &    (4)          &   (5)                  &   (6)        & (7)            &  (8)    \\
\hline
Io      &  15.0             & \,\,4  &  $10.3 \pm 8.4$ &$\rm 8.5\times 10^{-16}$&$\approx 0.6$ &  --            & ($ -2.00$) \\
Europa  &\,\,6.0            & 64     &\,\,$5.5\pm 3.5$ &$\rm 1.3\times 10^{-14}$&$\approx 1.0$ & $0.58 \pm 0.04$&$-2.02\pm 0.63$\\
Ganymede&\,\,8.2            & 38     &\,\,$7.2\pm 4.9$ &$\rm 9.5\times 10^{-16}$&$\approx 0.6$ & $0.82 \pm 0.04$&$-2.82\pm 2.60$\\
Callisto&\,\,7.2            & 35     &\,\,$6.4\pm 2.1$ &$\rm 5.2\times 10^{-16}$&$\approx 0.5$ & $0.86 \pm 0.14$&$-1.60\pm 0.39$\\
\hline
\hline
\end{tabular}
\end{table}

\begin{table}[tbh]
\caption
{
Mass budgets of the dust clouds (model estimates,
see Krivov et al. 2002b for description of the model).
}
\bigskip  
\begin{center}
\begin{tabular}{lcccc}
\hline  
\hline  
Satellite  &
Io         &
Europa     &
Ganymede   &
Callisto   \\
\hline
Mass flux of impactors [$\g\m^{-2}\second^{-1}$]&
$1 \times 10^{-11}$ &
$7 \times 10^{-12}$ &
$4 \times 10^{-12}$ &
$3 \times 10^{-12}$ \\
Mass inflow of impactors [$\g\second^{-1}$]&
130 &
 50 &
100 &
 50 \\
Yield&
$3 \times 10^4$ &
$2 \times 10^4$ &
$4 \times 10^3$ &
$7 \times 10^2$ \\
Mass production rate of ejecta [$\g\second^{-1}$]&
$4 \times 10^6$ &
$9 \times 10^5$ &
$4 \times 10^5$ &
$3 \times 10^4$ \\
Mean lifetime of ejecta [$\second$]$^{a)}$&
 50 &
 70 &
120 &
250 \\
Steady-state mass of the cloud [tons]&
200 &
 60 &
 40 &
  9 \\
Fraction of escaping ejecta&
$1 \times 10^{-4}$ &
$2 \times 10^{-4}$ &
$8 \times 10^{-4}$ &
$4 \times 10^{-3}$ \\
Ejection rate into circumjovian space [$\g\second^{-1}$]&
400 &
200 &
300 &
100 \\
\hline  
\hline
\end{tabular}

\vspace*{5mm}
$^{a)}$~Residence time within the Hill sphere before recollision or escape
\end{center}
\label{tab_budget}
\end{table}

%=========================================================================
\end{document}